\documentclass[%
reprint,
superscriptaddress,
amsmath,amssymb,
 prb,aps]{revtex4-2}

\usepackage{graphicx}
\usepackage{bm}
\usepackage{natbib}
\usepackage[caption=false]{subfig}

\usepackage{xcolor, soul}
\sethlcolor{yellow}

\begin{document}

\title{Transport scaling and critical tilt effects in disordered two-dimensional Dirac fermions}

\author{Swadeepan Nanda}
\affiliation{Department of Physics and Texas Center for Superconductivity, University of Houston, Houston, TX 77204, USA}
\author{Pavan Hosur}
\affiliation{Department of Physics and Texas Center for Superconductivity, University of Houston, Houston, TX 77204, USA}
\date{\today}
\begin{abstract}
Two-dimensional (2D) Dirac fermions occur ubiquitously in condensed matter systems from topological phases to quantum critical points. Since the advent of topological semimetals, where the dispersion is often tilted around the band crossing where the Dirac fermion can appear, tilt has emerged as a key handle that controls physical properties. We study how tilt affects the transport and spectral properties of tilted 2D Dirac fermions under scalar disorder. Although our spectral analyses always show conformity to appropriate Gaussian ensembles, suggestive of delocalization, the conductivity scaling $g(L)$ shows a surprising richness. For a single Dirac node, relevant for quantum Hall transitions and topological insulator surface states, we find $g(L)\sim a_1\log(L)$ with a tilt-dependent coefficient $a_1>0$. Interestingly, when the tilt and transport directions are aligned, $a_1$ and hence $g(L)$ shows a spike at the critical point between the type-I and type-II regimes of the Dirac node. For systems with two Dirac nodes with unbroken time-reversal symmetry, pertinent to quasi-2D Dirac materials, we find $g(L)\sim L^{a_1}(\log L)^{a_2}$. However, we find a surprising tension between tilt along and perpendicular to the transport directions. For the former, $a_1$ changes sign as a function of tilt, hinting at a tilt-driven localization-delocalization transition, while $a_1<0$ for all tilts in the latter case, implying localization. These localized behaviors also reveal tension with the delocalization seen in spectral properties and suggest differing localization tendencies in real and Hilbert spaces. Overall, our work identifies tilt as an essential control parameter that uncovers rich and unconventional transport physics in 2D Dirac materials.

\end{abstract}

\maketitle

\section{Introduction}
The phenomenon of localization, originally introduced by Anderson in 1958, has laid the foundation for the modern theoretical framework describing wave propagation in disordered media \citep{PhysRev.109.1492}. Anderson localization arises due to coherent quantum interference among multiple scattering paths in a disordered potential, leading to the suppression of diffusion and the exponential localization of eigenstates in real space. While initially formulated for non-interacting electrons in random lattices \citep{PhysRev.109.1492, PhysRevLett.133.066101}, the concept has since been extended to a wide variety of classical and quantum systems, including photonic structures \citep{wiersma1997localization, PhysRevB.96.144201,PhysRevA.110.053517}, acoustic media \citep{hu2008localization}, and ultracold atomic gases \citep{billy2008direct, jendrzejewski2012three, white2020observation, PhysRevResearch.6.033039, PhysRevLett.95.020401, PhysRevB.102.134206}. The likelihood and nature of localization are strongly influenced by both the dimensionality of the system and its underlying symmetries \citep{PhysRevB.40.5325}. Mechanistically, localization is driven by enhanced backscattering \citep{boguslawski2017observation}, phase coherence, and the absence of other delocalizing mechanisms \citep{RevModPhys.57.287, RevModPhys.80.1355}. In contrast, delocalization can emerge from topological protection \citep{RevModPhys.82.3045, RevModPhys.80.1355}, symmetry constraints \citep{gade1991n, PhysRevB.59.13221}, or dynamical effects \citep{PhysRevLett.114.056801,ponte2015periodically, PhysRevLett.95.206603, jain2017higgs} that suppress interference. Theoretical tools such as nonlinear sigma models \citep{pruisken1984localization, wegner1979mobility, PhysRevLett.45.1352}, random matrix theory \citep{RevModPhys.80.1355, RevModPhys.80.1355}, and perturbative renormalization group techniques \citep{Efetov:1997fw, PhysRevLett.42.673, PhysRevB.77.165108}, along with numerical methods including level spacing statistics, finite-size scaling, and multifractal analysis \citep{shklovskii1993statistics, mirlin2000multifractality}, have provided a robust framework for probing the critical behavior of localization–delocalization transitions across symmetry classes and dimensions. 

A fundamental route to understanding the effects of disorder in quantum systems is through the study of electrical conductivity, which provides a direct and experimentally measurable probe of the underlying quantum coherence and localization phenomena. The pioneering work of Anderson demonstrated that sufficiently strong disorder can localize electronic wavefunctions, thereby inhibiting charge transport even in the absence of interactions \citep{PhysRev.109.1492}. This insight marked a conceptual shift from semiclassical transport theory to one governed by quantum interference. In weakly disordered metals, electrons propagate diffusively, and conductivity remains finite, albeit corrected by weak-localization or weak-antilocalization effects depending on the system’s symmetry. However, as disorder strength increases, interference between multiply scattered electronic paths enhances backscattering, eventually suppressing diffusion and driving a metal–insulator transition.
The subsequent formulation of the scaling theory of localization by Abrahams et al. \citep{PhysRevLett.42.673} provided a unifying description of these phenomena by relating the dimensionless conductance $g=G/(e^2/h)$ to the system size $L$. The scaling function $\beta(g)=\frac{d\log g}{d \log L}$ captures the flow of conductivity with length scale: $\beta(g)>0$ corresponds to metallic behavior where conductance increases with system size, $\beta(g)<0$ signals insulating behavior with decaying conductance, and $\beta(g)=0$ identifies a quantum critical point separating these regimes.

While transport probes the macroscopic manifestations of localization through the scaling of conductivity \citep{Abrahams1979}, the spectral properties of the underlying Hamiltonian reveal its microscopic origin in the spatial structure of eigenstates. In particular, the statistical analysis of energy levels provides a complementary framework to distinguish localized, metallic, and critical regimes based on correlations between neighboring eigenvalues \citep{Shklovskii1993, Kravtsov2020}. This approach, rooted in random matrix theory \citep{Mehta2004, Haake2010}, provides deep insights into the underlying quantum dynamics: whether the system exhibits localized, metallic (ergodic), or critical (non-ergodic) behavior. In particular, level statistics have emerged as a key diagnostic in studies of Anderson localization, quantum chaos, and disordered topological systems \citep{Evers2008, Altland1997, Chalker2010}. In the absence of disorder, the energy levels of a clean periodic system follow a regular and predictable pattern, reflecting the translational symmetry of the underlying lattice. However, in a disordered system, this symmetry is broken, and the eigenvalue distribution becomes nontrivial. Importantly, the nature of the eigenstates, whether they are extended or localized, directly influences the statistical correlations between neighboring energy levels \citep{Oganesyan2007, Atas2013}.

When eigenstates are strongly localized, they reside on spatially distinct regions of the lattice and do not interact with each other. As a result, their energy levels are uncorrelated, and the level spacing distribution follows a Poisson distribution:
$P_{\text{Poissonian}}(s) = e^{-s}$, where $s$ is the spacing between adjacent energy levels normalized by the local mean level spacing. This lack of level repulsion reflects the absence of hybridization between spatially disconnected states. In contrast, when states are delocalized and ergodic, they overlap and hybridize strongly, producing level repulsion. In this regime, level statistics follow the Wigner-Dyson distributions, classified according to the symmetries of the system~\citep{mehta1963statistical, zirnbauer1996riemannian}: Gaussian Orthogonal Ensemble (GOE): applies when time-reversal symmetry is preserved and spin-rotation symmetry is unbroken (e.g., spinless or spinful systems without spin-orbit coupling). The level spacing distribution is:
$P_{\text{GOE}}(s) = \frac{\pi}{2}se^{-\frac{\pi}{4}s^2}$. Gaussian Unitary Ensemble (GUE): applies when time-reversal symmetry is broken, e.g., by applying a magnetic field or including complex hopping amplitudes. The distribution becomes $P_{\text{GUE}}(s) = \frac{32}{\pi^2}s^2e^{-\frac{4}{\pi}s^2}$ Gaussian Symplectic Ensemble (GSE): relevant when TRS is preserved but spin-rotation symmetry is broken, typically due to strong spin-orbit coupling. This leads to  $P_{\text{GSE}}(s) = \frac{2^{18}}{3^6\pi^3}s^4e^{-\frac{\pi}{4}s^2}$.

It is well established that disordered systems in one or two dimensions localize for arbitrarily weak disorder ~\citep{mott1961theory,ishii1973localization,gol1977pure,kramer1993localization} as a consequence of restricted phase space and enhanced quantum interference effects. However, this conventional picture changes in the presence of spin–orbit coupling (SOC). In 2D systems with strong SOC, interference between time-reversed paths acquires an additional phase due to spin rotation, leading to destructive interference of backscattered waves. This mechanism gives rise to weak antilocalization, characterized by enhanced conductivity at low temperatures and a metallic response for weak disorder ~\citep{hikami1980,bergmann1984}.
A particularly striking realization of this behavior occurs in 2D Dirac systems, which may be viewed as the extreme limit of SOC-driven antilocalization. The Dirac Hamiltonian intrinsically locks spin (or pseudospin) to momentum, ensuring that backscattering between time-reversed states is strongly suppressed. Early theoretical studies have reported that certain classes of disordered 2D Dirac semimetals do not conform to the expectations of conventional localization theory \citep{PhysRevLett.98.256801,PhysRevLett.99.146806,PhysRevLett.99.106801}. Specifically, when disorder preserves key symmetries such as chiral, sublattice, or time-reversal symmetry, the system may remain at a critical point or exhibit marginal delocalization, even in the presence of significant disorder.
Several mechanisms have been proposed to account for this anomalous behavior. Firstly, the linear dispersion of Dirac fermions yields a vanishing density of states at the nodal point, suppressing the phase space available for scattering and thereby weakening localization tendencies. Secondly, the $\pi$ Berry phase associated with closed trajectories around Dirac points leads to destructive interference of backscattered paths, diminishing quantum localization effects. Thirdly, the topological protection of Dirac nodes and the associated stability of zero modes under certain disorder types can preserve extended or critical states. Together, these features place 2D Dirac semimetals in a unique position at the boundary between conventional and anomalous localization, motivating ongoing theoretical and numerical efforts to delineate the precise conditions under which delocalization persists.

In this work, we investigate how the localization behavior of 2D Dirac semimetals is modified when one of the key low energy constraints is relaxed. Specifically, we focus on the effect of tilt in the Dirac dispersion - a perturbation that breaks the emergent Lorentz invariance of the low energy theory by introducing anisotropy and a preferred direction in momentum space. While tilt does not by itself alter the fundamental symmetry class in a multi node description, in a single node effective theory it breaks time-reversal ($\mathcal{T}$) symmetry~\citep{moradpouri2024kinetic}. In the upright case, 2D Dirac semimetals preserve effective chiral or $\mathcal{T}$ symmetries, which suppress localization by enabling destructive interference and topological protection \citep{nomura2006quantum, groth2009theory}. The introduction of a finite tilt not only induces anisotropic quasiparticle velocities but also modifies the density of states (DOS) near the Dirac point and disrupts the interference structure. As a result, the symmetry class of the system changes, and the disorder flows differently under renormalization \citep{mirlin2000multifractality}. Tilt does not directly enhance localization but rather accelerates the transition from a critical phase to a disordered metallic regime. It has been shown in \citep{yang2018effects} that for a tilted cone, even arbitrarily weak potential disorder causes the Dirac point to acquire a finite DOS and finite scattering rate at low energy. The clean band-touching point does not survive; instead, it is replaced (in a renormalized, disorder-averaged sense) by an extended region of states sometimes described as a “bulk Fermi arc” \citep{PhysRevB.100.125138, PhysRevB.98.195123} in the Brillouin zone. Thus, a tilted Dirac semimetal is inherently more unstable to becoming a metal. In fact, the extreme case where the cone is over-tilted, dubbed Type II, the clean system is already metallic with finite DOS at the zero energy due to band overlap, so adding disorder simply enhances the diffusive character further. 

From an experimental standpoint, the study of tilt in Dirac materials is strongly motivated by its ubiquitous presence in realistic systems and its profound impact on electronic properties~\citep{armitage2018weyl}. Tilted Dirac cones have been observed in a wide range of materials, including strained graphene \citep{pereira2009strain}, borophene~\citep{islam2017signature, PhysRevB.94.165403, lopez2016electronic, verma2017effect}, organic conductors such as $\alpha$-(BEDT-TTF)$_2$I$_3$ \citep{kobayashi2007massless, PhysRevB.77.115117}, thin films ~\citep{rizza2022extreme} and transition metal dichalcogenides ~\citep{PhysRevX.6.031021,chang2016prediction}, where the tilt arises due to lattice strain, substrate-induced symmetry breaking, or electronic correlations \citep{katayama2006pressure, goerbig2008tilted, dressel2004optical}. In Weyl semimetals, tilt serves as a key distinguishing feature between type-I and type-II phases, with the latter characterized by a tipping over of the Weyl cones and the emergence of open Fermi surfaces \citep{soluyanov2015type, chang2016prediction, yan2017topological}. Importantly, the degree of tilt is often tunable via external control parameters such as uniaxial strain, hydrostatic pressure, or chemical doping \citep{nishine2010anisotropy, xi2015strongly, su2018strain}, enabling dynamic control over the anisotropy of the band structure and associated transport signatures. This tunability positions tilted Dirac and Weyl systems as promising candidates for strain-engineered quantum devices, where the manipulation of band tilt could enable new regimes of anisotropic conductivity, tunable chiral anomaly, or magnetotransport effects \citep{zhao2015anomalous, trescher2015quantum, galeski2021unusual}. From a transport perspective, tilt reshapes the DOS near the Dirac point, modifies the group velocity landscape, and redistributes the available phase space for scattering. These factors collectively influence conductivity scaling, suppress weak antilocalization, and can even induce a crossover from critical to diffusive behavior in disordered systems \citep{ferreiros2017anisotropic, narayanan2015quasiparticle, burkov2014anomalous}.

Taken together, these developments underscore that understanding the role of tilt is essential not only for interpreting electronic and transport measurements in realistic materials, but also for exploring new quantum phases at the intersection of disorder, topology, and non-Hermiticity~\citep{PhysRevB.99.201107,kozii2024non, zyuzin2018flat}. Tilted Dirac systems thus offer a rich platform for studying the interplay of geometry, symmetry, and localization in 2D and 3D electronic systems.

\section{RESULTS}

\subsection{Setup}

In this work, we investigate the effects of tilt on 2D Dirac fermions in the presence of disorder. Our models consist of a one and two tilted Dirac nodes in presence of disorder potential. For single and double nodes tilted along $x$ axis, the clean Hamiltonians read: 
\begin{equation}
\begin{aligned}
& H_0 = 
&\begin{cases}
\boldsymbol{k}\!\cdot\!\boldsymbol{\sigma} + \lambda k_x, 
  & \text{single node}, \\[6pt]
(K_0^2 - k_x^2)(\sigma_x+\lambda) + k_y\sigma_y,\label{eq:2node}
  & \text{two nodes}.
\end{cases}
\end{aligned}
\end{equation}
where $\pm K_0$ indicates the locations of the nodes. The $\lambda k_x$ term of the single node Hamiltonian breaks $\mathcal{T}$ and yields the unitary (AI) class. On the other hand, the Hamiltonian of two nodes preserves $\mathcal{T}$ with $\mathcal{T}^2 = I$, and thus belongs to the orthogonal (A) class.

To probe the interplay between tilt and disorder, we introduce scalar potential disorder, modeled as a spatially uncorrelated, $\mathcal{T}$ symmetric potential disorder $V(\boldsymbol{r})$ arising from randomly distributed impurities located at positions $\boldsymbol{R_I}$. Each impurity contributes to the total disorder potential through a scattering function $U(\boldsymbol{r}-\boldsymbol{R_I})$, such that the overall disorder potential takes the form $V(\boldsymbol{r})=\sum_{I=1}^{N}U(\boldsymbol{r}-\boldsymbol{R_I})$. We assume a Gaussian scattering potential, which, in momentum space, takes the form $U(\boldsymbol{q})= ue^{-q^2l_0^2/2}$. Here $u$ denotes disorder strength, $l_0$ dictates the range of the potential, and $N$ is the number of impurity centers. With this setup, we study transport properties through conductivity scaling as well as spectral properties such as level statistics and the ratio of consecutive spacings (RCS). 

We compute the conductivity (which is equivalent to the conductance in 2D) of tilted disordered Dirac Hamiltonian $H = H_0 + V(\boldsymbol{r})$ by evaluating Kubo formula
\begin{equation}
    g_{ii} = \frac{i2\pi\hbar^2}{L^2}\sum_{n,m}\frac{f(E_n) - f(E_m)}{E_n - E_m}\frac{\langle n|v_i|m\rangle \langle m|v_i|n\rangle }{E_n - E_m + i\eta}
    \label{eq: kubo}
\end{equation}
where $\boldsymbol{v} = \frac{\partial H}{\partial \boldsymbol{k}}$, $f(E)$ is the Fermi-Dirac distribution at zero temperature, $\eta$ is a small broadening parameter that accounts for the finite switching time of the applied field, $|n\rangle$ is an eigenstate of the random Hamiltonian $H$ with energy $E_n$. Our computation follows the approach in Refs. \citep{PhysRevLett.99.146806, PhysRevLett.99.106801, PhysRevLett.98.076602}.

The Hamiltonian is implemented in a momentum-pseudospin basis with a circular cutoff $\Lambda = 20\times\frac{2\pi}{L}$. The disorder potential is a sum over randomly distributed impurity centers, each contributing a Gaussian scattering potential, which, in momentum space, becomes
\begin{equation}
 \langle \boldsymbol{k}\sigma | V | \boldsymbol{k}'\sigma'\rangle = \frac{1}{L^2}\sum_{I = 1}^{N}U(\boldsymbol{k}-\boldsymbol{k'})e^{(\boldsymbol{k}-\boldsymbol{k}')\cdot\boldsymbol{R_I}}\delta_{\sigma, \sigma'}   
 \label{eq:disorder}
\end{equation}
The conductivity is averaged over a large ensemble of disorder realizations, and the errorbars shown in the plots are indicative of this disorder averaging(typically $1000$). 

The choice of $\eta$ plays a crucial role in ensuring physical reliability of the conductivity. One physically motivated choice sets $\eta \sim \frac{\hbar}{T_L}$, where $T_L$ is the escape time inferred from the Thouless energy, the spectral shift between periodic and antiperiodic boundary conditions. Alternatively, $\eta$ can be estimated as the inverse of the DOS at the Fermi energy, $\eta \sim 1/\rho(E=0)$, providing a dynamical scale based on contributing states to transport at relevant energy scale. Some other representations have also been found in literature i.e., Ref. \citep{PhysRevLett.98.076602} computed the conductivity for a range of $\eta$ values and selected the value that maximizes the conductivity. Here, for $\eta$, we chose a small energy window $E_{\text{cut-off}}$ across zero and took $\eta$ to be the inverse of number of states within $E_{\text{cut-off}}$.

\subsection{Single node}
\begin{figure}
  \centering

    \subfloat[]{%
        \includegraphics[width=0.3\linewidth]{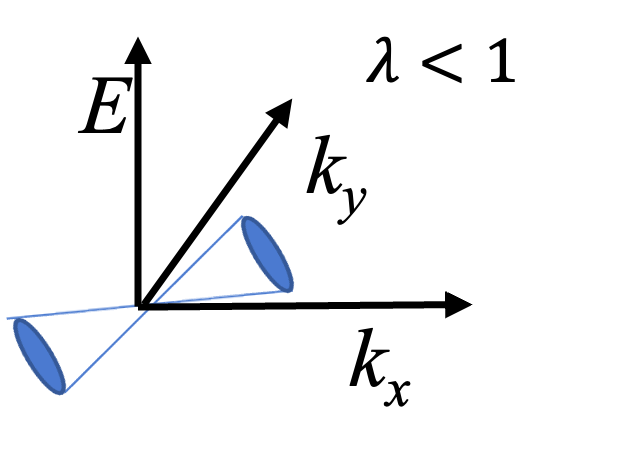}
        \label{fig:singlet1}%
    }
    \hfill
    \subfloat[]{%
        \includegraphics[width=0.3\linewidth]{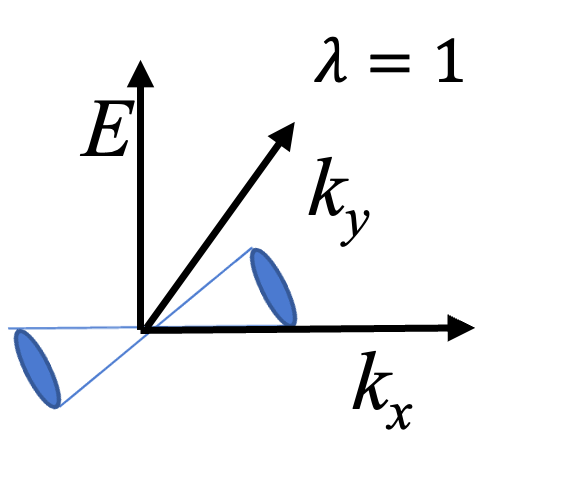}%
        \label{fig:singletc}}%
    \hfill
    \subfloat[]{%
        \includegraphics[width=0.3\linewidth]{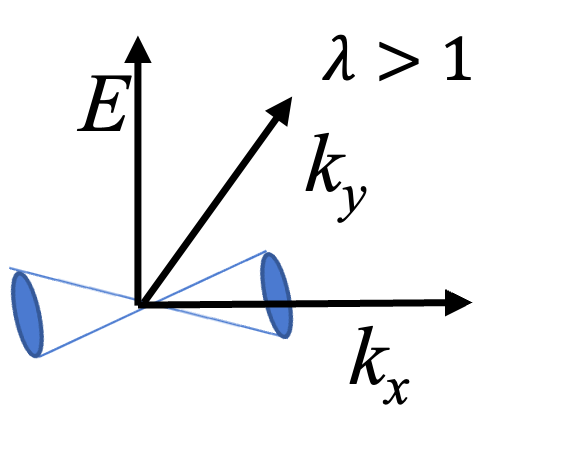}%
        \label{fig:singlet2}%
    }

\caption{Figs. \ref{fig:singlet1}, \ref{fig:singletc}, \ref{fig:singlet2} are showing type I, critical, and type II single Dirac node respectively.}
\label{fig:single}

\end{figure}

A single 2D Dirac fermion can arise in several distinct physical settings. The most prominent realization is on the surface of a three-dimensional topological insulator, where $\mathcal{T}$ symmetry protects the node against localization. It can also emerge effectively in graphene when intervalley scattering between the two different valleys is negligible, allowing each valley to behave as an independent Dirac species. Another realization occurs at the quantum Hall plateau transition, where a single Dirac cone describes the critical delocalized state between two topologically distinct insulating phases. In this section, we will discuss both transport and spectral properties as a function of tilt of the Dirac node, shown schematically in Fig. \ref{fig:single}.

\subsubsection{DOS}
 The DOS plays an important role in understanding the electronic properties of disordered systems, particularly in relation to localization phenomena. For clean Dirac fermion, the DOS vanishes linearly with energy for 
upright Dirac fermions.
On the other hand, tilt induces anisotropy in the energy spectrum and the constant energy contours deform from circles into elongated, direction-dependent shapes, and the group velocity becomes asymmetric across momentum directions. These changes have two key effects on the DOS. First, the elongation of the energy contours in certain directions increases the number of momentum states lying on a fixed-energy surface. Second, the group velocity is suppressed in regions along the tilt, which amplifies the contribution of those momentum states to the DOS. Together, these effects enhance the DOS at low energy compared to the upright case. As the tilt increases, the enhancement grows, and in the extreme limit of a type-II Dirac node, the DOS remains finite at zero energy due to the presence of overlapping electron and hole pockets. Explicitly, 
\begin{equation}
\begin{aligned}
& \rho(E) = 
&\begin{cases}
\frac{2|E|}{\pi(1-\lambda^2)^{3/2}}, 
  & \lambda<1, \\[6pt]
\frac{\Lambda}{\pi^2(\lambda^2-1)},
  & \text{at } E=0,\text{ and }\lambda>1
\end{cases}
\end{aligned}
\label{eq:dos}
\end{equation}
\begin{figure}
    \includegraphics[width=0.9\linewidth]{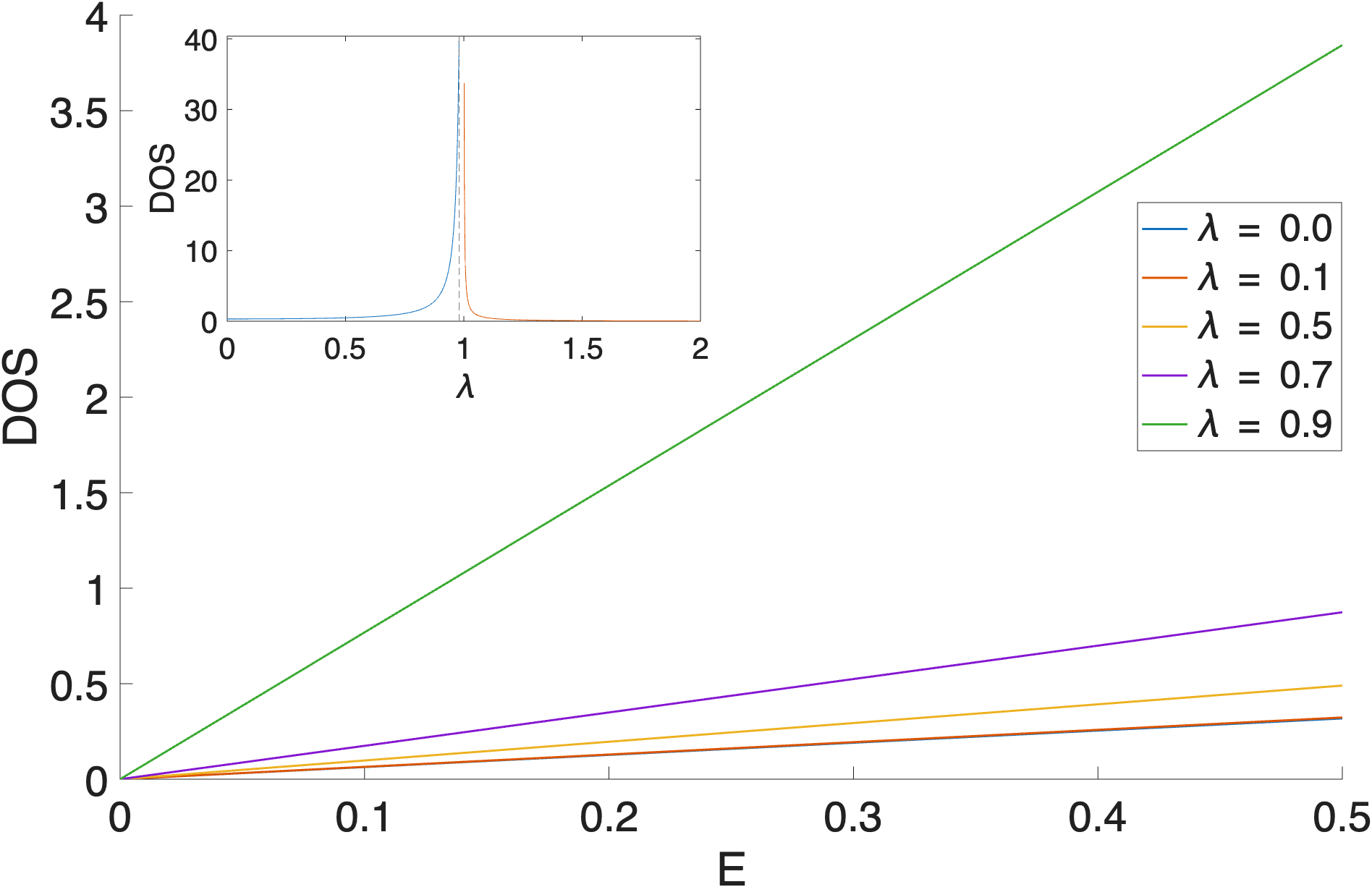}
    \caption{DOS of clean Dirac fermion for different tilt parameter $\lambda$. For $\lambda = 0$ the DOS is the lowest and it continues to grow as the nodes are more and more tilted until the nodes are critically tilted.}
    \label{fig:ldos}
\end{figure}
See Appendix \ref{app:dos} for detailed derivations. Note the divergence at $|\lambda|=1$, which will play a key role in the behavior of the conductivity that follows.

\subsubsection{Conductivity}

\begin{figure}[ht]
    \centering

    \subfloat[]{%
        \includegraphics[width=0.9\linewidth]{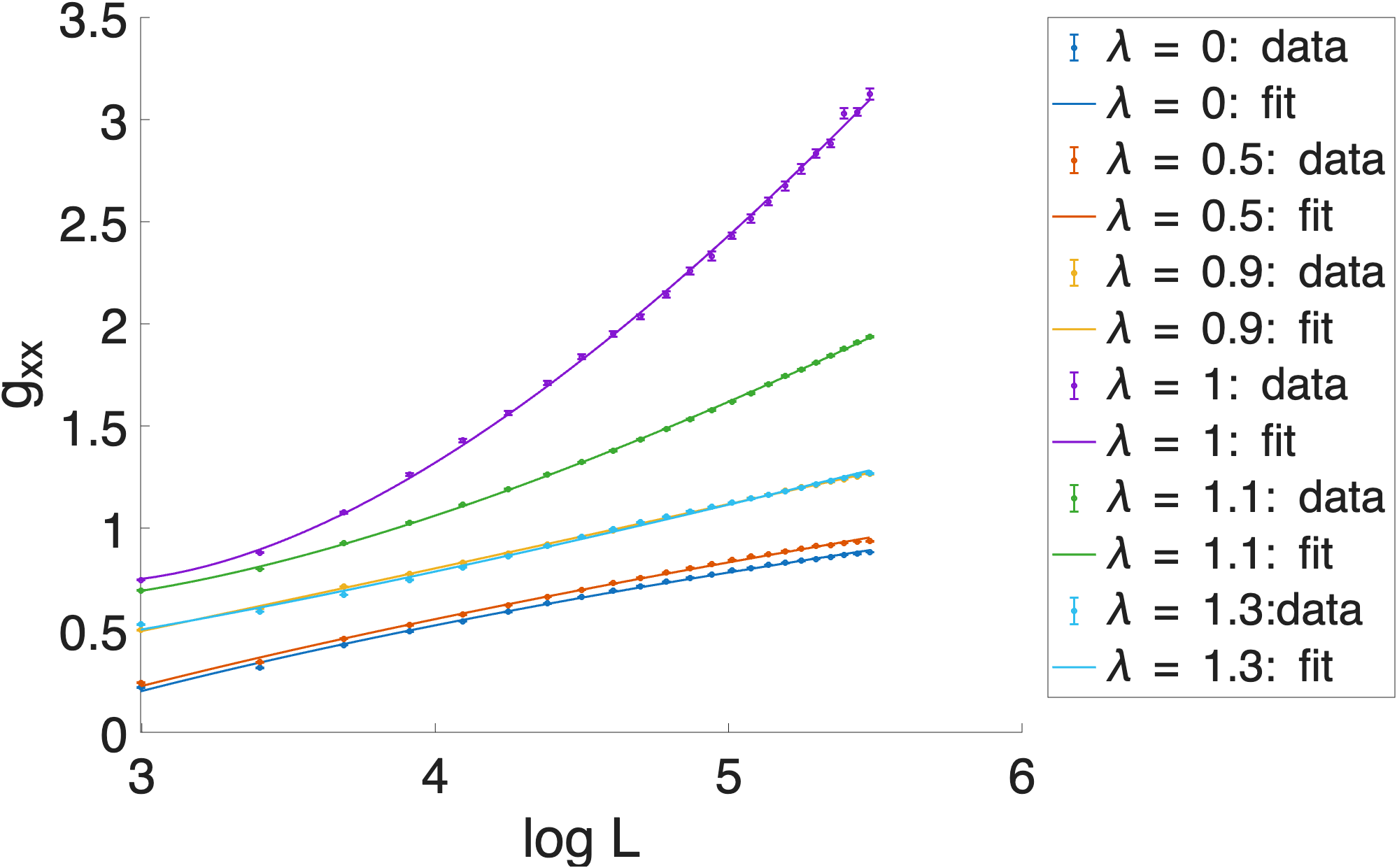}
        \label{fig:gxx}%
    }
    \hfill
    \subfloat[]{%
        \includegraphics[width=0.9\linewidth]{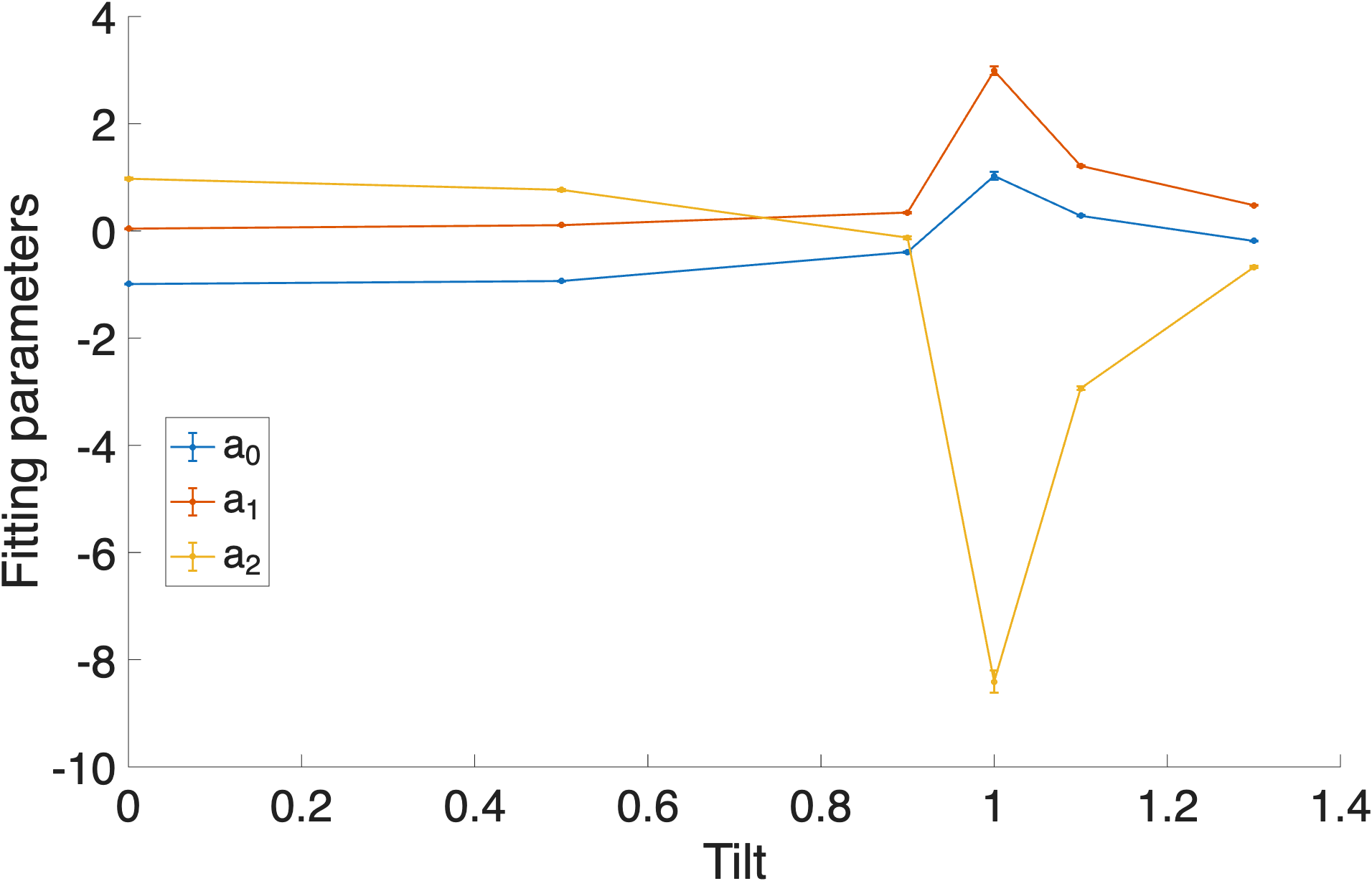}%
        \label{fig:abxx}%
    }

    \caption{Conductivity of a single Dirac node along the $x$ direction, $g_{xx}$, for different $\lambda$. For each $\lambda$, the data was obtained by averaging over around 1000 configurations. The obtained data was fitted against $g_{xx}(L) = a_0 + a_1 \log L + a_2\log\log L$. $a_0$ and $a_1$ for different $\lambda$ have been plotted in~\ref{fig:abxx}. A spike in $a_1$ and hence the conductivity has been observed for $\lambda = 1$. Errorbars are due to disorder averaging.}
\end{figure}

\begin{figure}
  \centering

    \subfloat[]{%
        \includegraphics[width=0.9\linewidth]{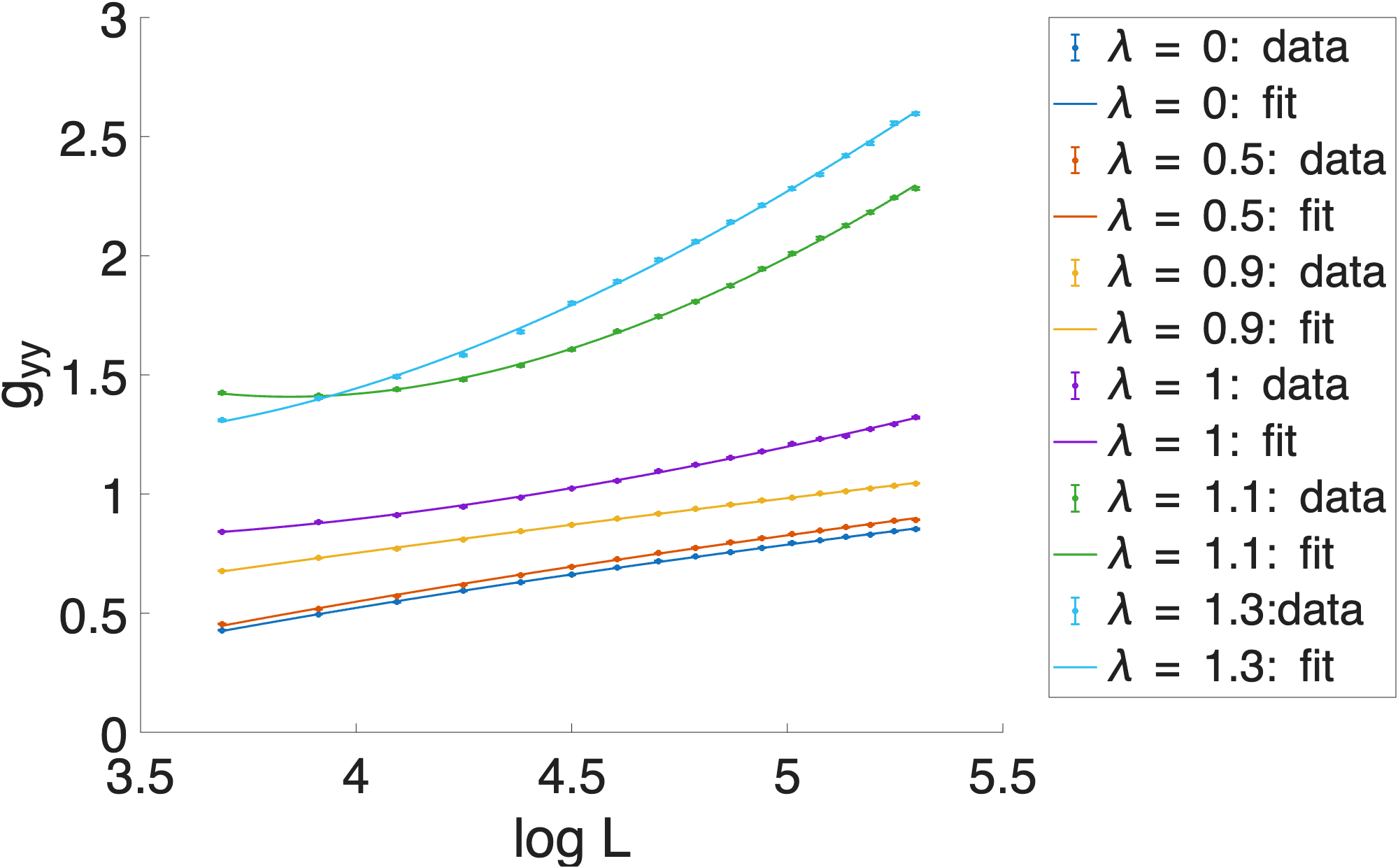}
        \label{fig:gyy}%
    }
    \hfill
    \subfloat[]{%
        \includegraphics[width=0.9\linewidth]{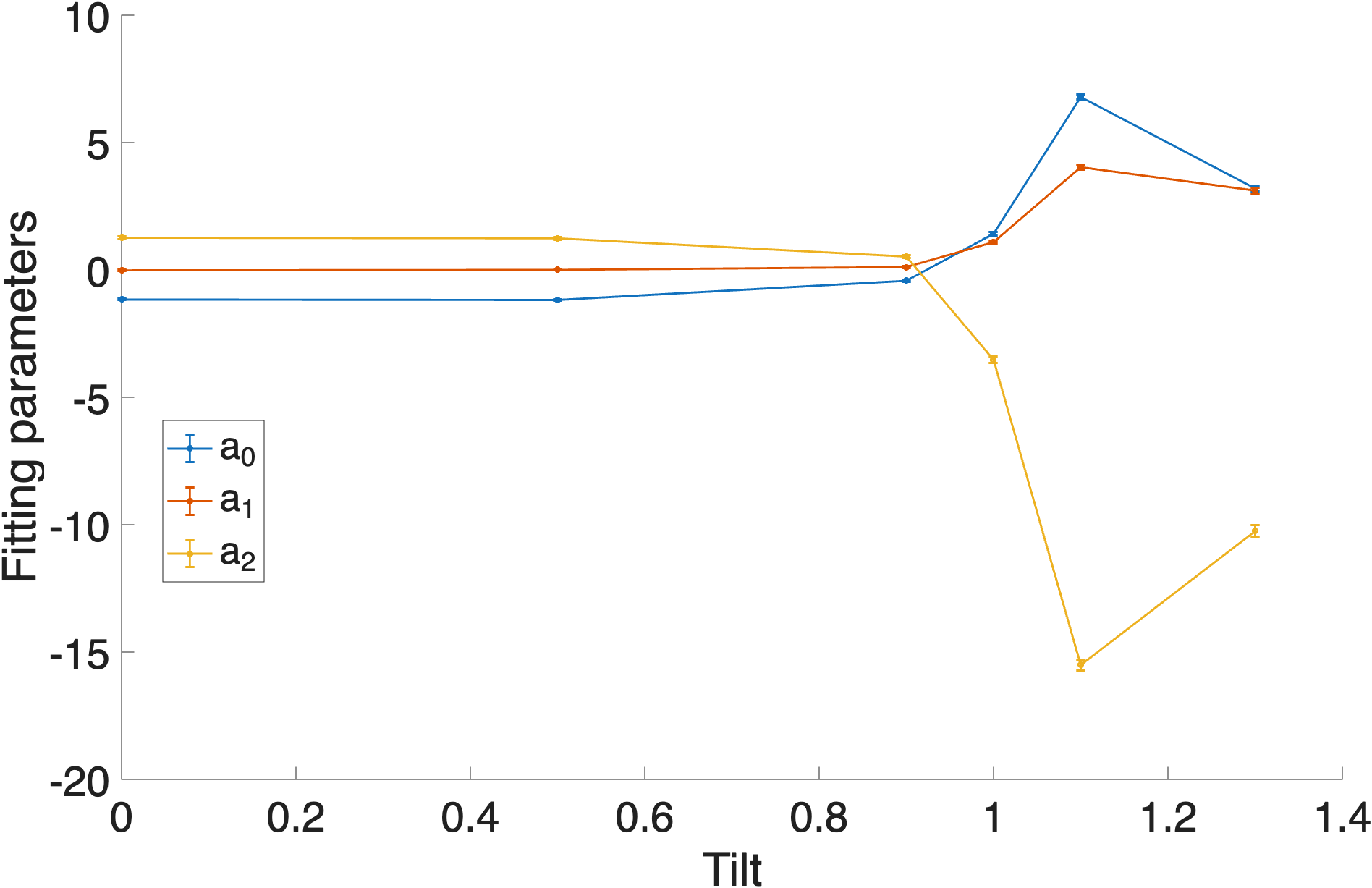}%
        \label{fig:abyy}%
    }
\caption{Conductivity of single Dirac node along $y$ direction $g_{yy}$ for different $\lambda$. For each $\lambda$, the data was obtained by averaging over around $1000$ configurations. The obtained data was fitted against $g_{yy}(L) = a_0 + a_1 \log{L} + a_2\log\log L$. $a_0$ and $a_1$ for different $\lambda$ have been plotted in~\ref{fig:abyy}. Unlike $g_{xx}$, monotonic behavior has been observed for $g_{yy}$, and $a_1$ (most important in the thermodynamic limit) continues to grow with $\lambda$}
\end{figure}

Starting from Eq. (\ref{eq: kubo}), using the identity $ \frac{1}{E_n-E_m+i\eta}=\mathcal{P}\frac{1}{E_n-E_m} - i\pi\delta(E_n-E_m)$, where $\mathcal{P}$  indicates the principal value, and after some manipulations (See Appendix \ref{app:analytics} for details), we arrive at 
\begin{equation}
    g_{xx}=\int dE_n\left(\frac{-\partial f}{\partial E_n}\right)\rho(E_n)\langle(v_x^n)^2\tau_n^{\text{tr}}\rangle_{E_n}
    \label{eq:kubotr}
\end{equation}
where $\rho(E_n)$ is the DOS at $E_n$, $E_n$ is the eigenvalue of $H$ corresponding to $n$-th level,  $\tau_{\text{tr}}^n$ is the transport lifetime of the $n$-th level, $v_x^n=\langle n|\frac{\partial H}{\partial k_x}|n\rangle$.  Eq. (\ref{eq:kubotr}) contains two competing factors as we vary tilt: $\rho(E)$ and $v_x^2$. Specifically, at $\lambda = 1$, from Eq. (\ref{eq:dos}), we notice that $\rho(E)$ diverges, and, in the weak disorder limit, from Eq. (\ref{eq:v}), $v_{x,s}^2$ vanishes for $s=\pm$ at $\phi=\pi,0$. To see what dominates, we study how conductivity scales with system size as a function of tilt in Fig. \ref{fig:gxx} $(g_{xx})$ and Fig. \ref{fig:gyy} $(g_{yy})$. For transport along the tilt direction, we observe some interesting pattern with tilt. It is the lowest for an upright node,  and increases with $\lambda$. If we look at Fig. \ref{fig:gxx} closely, we notice that change in $g_{xx}$ is little for $\lambda\lesssim1$. However, we observe sudden large jump in $g_{xx}$ at $\lambda=1$ when the node is critically tilted. Upon further tilting i.e., for $\lambda>1$ when the node becomes type 2, $g_{xx}$ starts to fall. Thus, $g_{xx}$ is peaked around $\lambda=1$. In Appendix \ref{app:analytics}, we show analytically that for small tilt, $\lambda\ll1$, $g_{xx}$ increases with $\lambda$, and the change occurs at $O(\lambda^2)$, while for $\lambda>1$, $g_{xx}$ monotonically decreases. We can also infer this behavior from Fig. \ref{fig:ldos}, where the DOS is also peaked at $\lambda=1$.  However, the scenario is different for $g_{yy}$ -- it does not show a maximum at $\lambda = 1$, but grows monotonically with $\lambda$ at a rate that increases as we enter the type-II regime.

The data in Fig. \ref{fig:gxx} and Fig. \ref{fig:gyy} shows that conductivity fits well to
\begin{equation}
 g=a_0+a_1\log L   
\end{equation} or equivalently, 
\begin{equation}
  \beta(g)=a_1/g   
 \end{equation}
 From Fig. \ref{fig:abxx} and Fig. \ref{fig:abyy}, we can infer conductivity at larger length scales. For $L\to\infty$, the log term will dominate and the coefficient $a_1$ will dictate the behavior. Thus, there will be spike in $g_{xx}$ at the Lifshitz transition, $\lambda=1$, while $g_{yy}$ will grow monotonically with $\lambda$ at a rate that is higher in the type-II regime.
Overall, since $g(L)>0, \beta > 0$, a single Dirac node always remains delocalized irrespective of tilt and disorder strength, consistent with the prediction of \citep{PhysRevLett.99.146806} for an upright Dirac node being a special case of our result.

The argument in Ref. \citep{PhysRevLett.99.146806} pertains to time-reversal-symmetric Dirac fermions on the surface of a three dimensional topological insulator, where delocalization is protected by the $Z_2$ topology of the bulk. In contrast, tilted Dirac fermions explicitly break $\mathcal{T}$ symmetry and can realize the critical point between two quantum Hall insulating phases, known as the quantum Hall plateau transition \citep{Pruisken1988,Huckestein1995,Evers2008}. Despite the distinct physical origins, both systems exhibit similar scaling behavior of conductivity, indicating that such delocalization is an intrinsic feature of 2D Dirac fermions. Both realizations are topological, but in fundamentally different ways: in the former, delocalization arises from $Z_2$ topological protection, whereas in the latter it is ensured by criticality between two topological phases.

\subsubsection{Level statistics}
In disordered Dirac systems, level spacing statistics are widely used to classify universality classes and examine the presence or absence of spectral correlations~\citep{AltlandZirnbauer1997,EversMirlin2008,Shklovskii1993}. To examine the universality class, we compute the exponent $\alpha$ by fitting the distribution of energy spacings $P(s)$ between nearest energy levels against $P(s)= As^{\alpha}e^{-Bs^2}$. This functional form originates from random matrix theory and captures key features of spectral correlation in disordered systems: the power law prefactor $(s^{\alpha})$ indicates level repulsion for small spacing $s$, the exponential part $(e^{-Bs^2})$ represents Gaussian suppression of large spacings ~\citep{Mehta2004,Haake2010}, and $A,B$ are obtained from
\begin{equation}
    \int_{0}^{\infty}P(s)ds = 1,\qquad \int_{0}^{\infty}sP(s)ds = 1
\end{equation}

\begin{figure}

\centering

    \subfloat[]{%
        \includegraphics[width=0.9\linewidth]{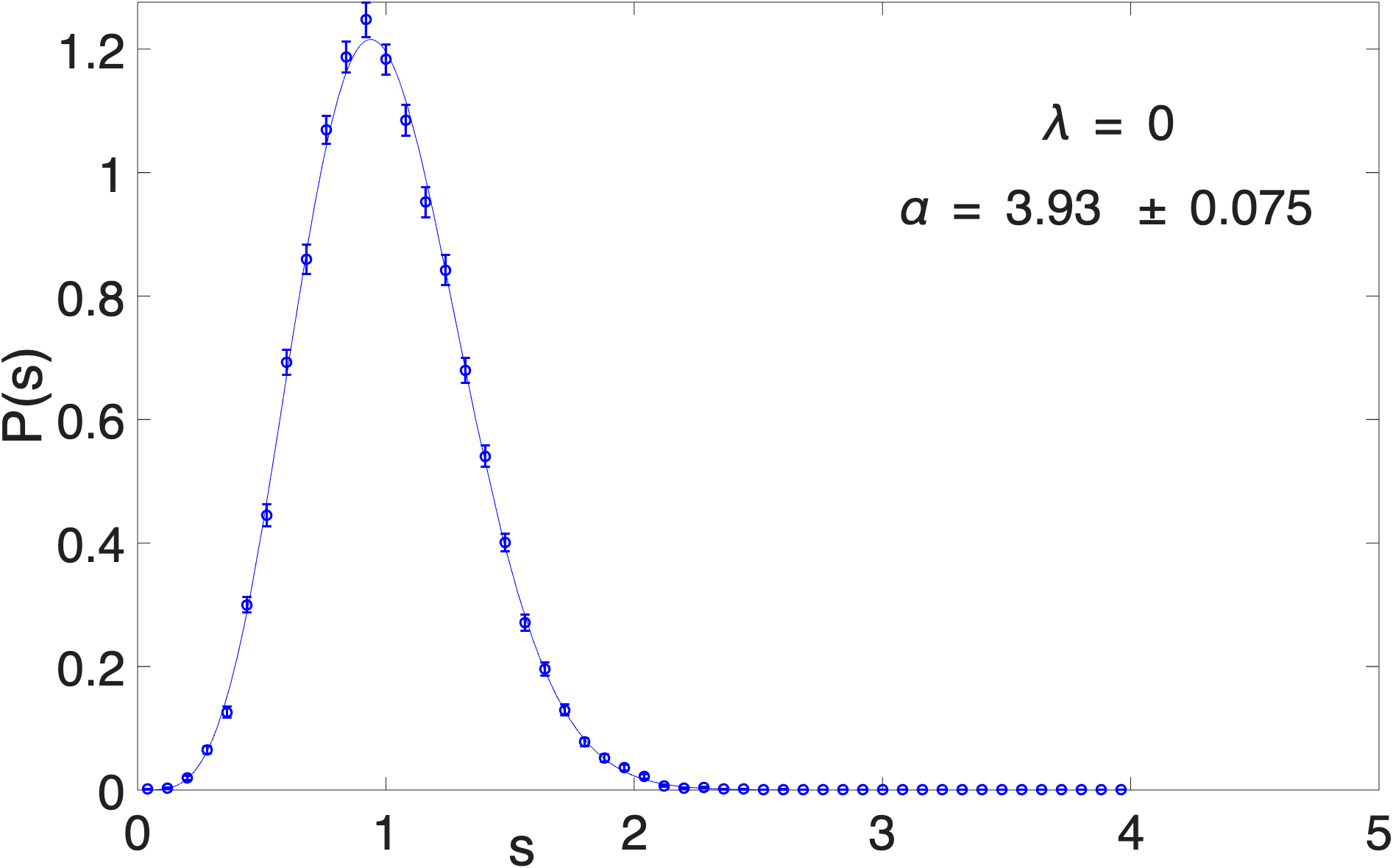}
        \label{fig:ls1n}%
    }
    \hfill
    \subfloat[]{%
        \includegraphics[width=0.9\linewidth]{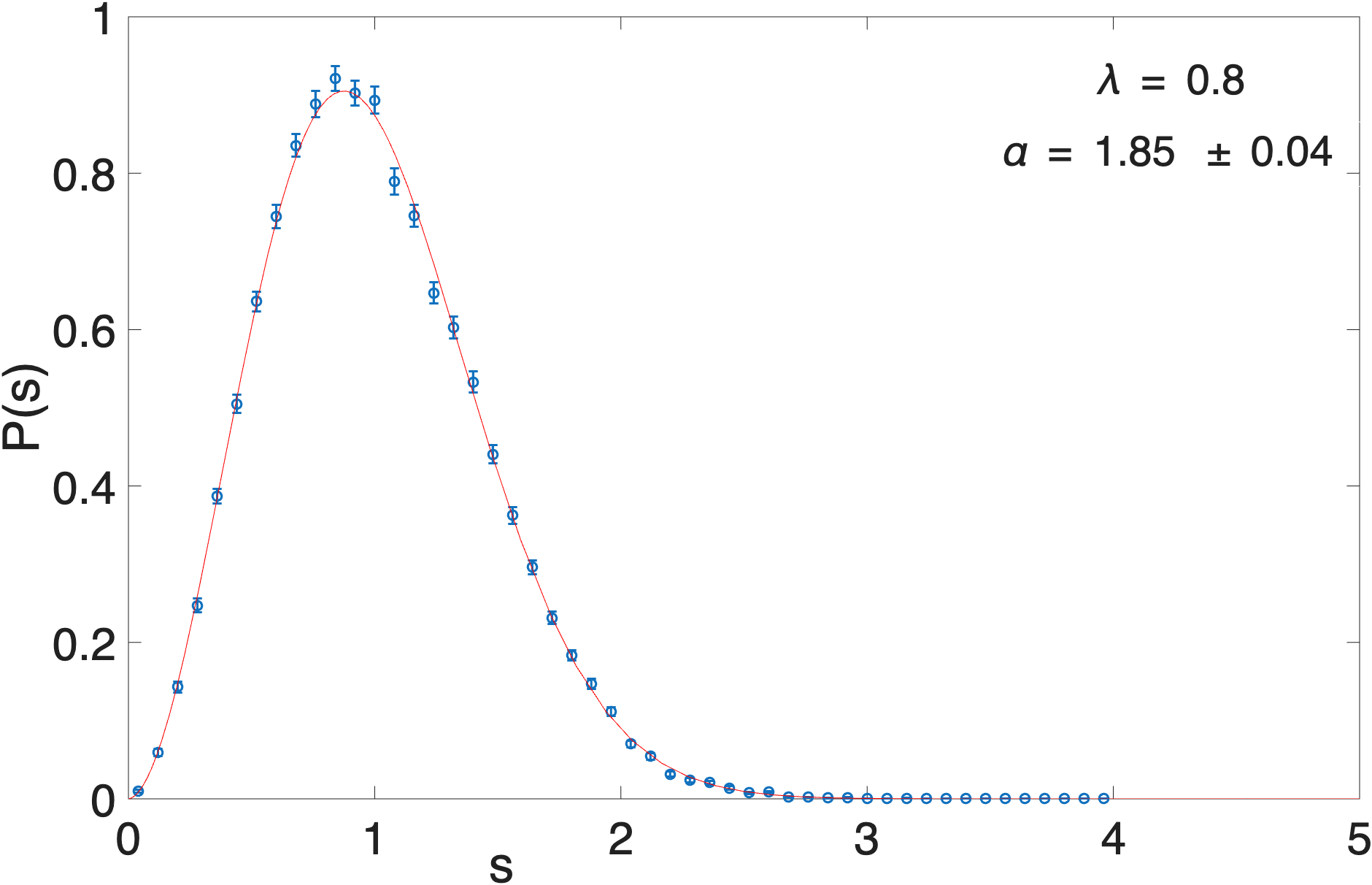}%
        \label{fig:ls1t}%
    }
\caption{Level spacing distribution of a single Dirac fermion with $\lambda = 0, 0.8$ in \ref{fig:ls1n} and \ref{fig:ls1t} respectively. The data has been fitted against $f(s) = As^{\alpha}e^{-Bs^2}$, where $A, B$ are determined from normalization. For the tilted Dirac systems with single node, we obtain a fitted value of $\alpha \approx 1.85$, which is close to the expected value of $\alpha = 2$ for the GUE. In contrast, for the upright Dirac fermion, the fitting yields $\alpha \sim 3.93$, in good agreement with $\alpha = 4$ associated with the GSE. }
\end{figure}

For an upright Dirac fermion with scalar disorder, one expects the system to lie in the symplectic class (GSE) due to the presence of time-reversal symmetry and strong spin-orbit coupling, which is characterized by with $\alpha=4$. Our level spacing analysis reveals $\alpha \sim 3.93$, in good agreement with $\alpha=4$. Introducing a tilt to the Dirac cone explicitly breaks time-reversal symmetry, formally pushing the system into the GUE. Here, in Fig. \ref{fig:ls1n}, we recover $\alpha \sim 1.85$ which is also close to $\alpha = 2$ for GUE.

\begin{figure}
\centering

    \subfloat[]{%
        \includegraphics[width=0.9\linewidth]{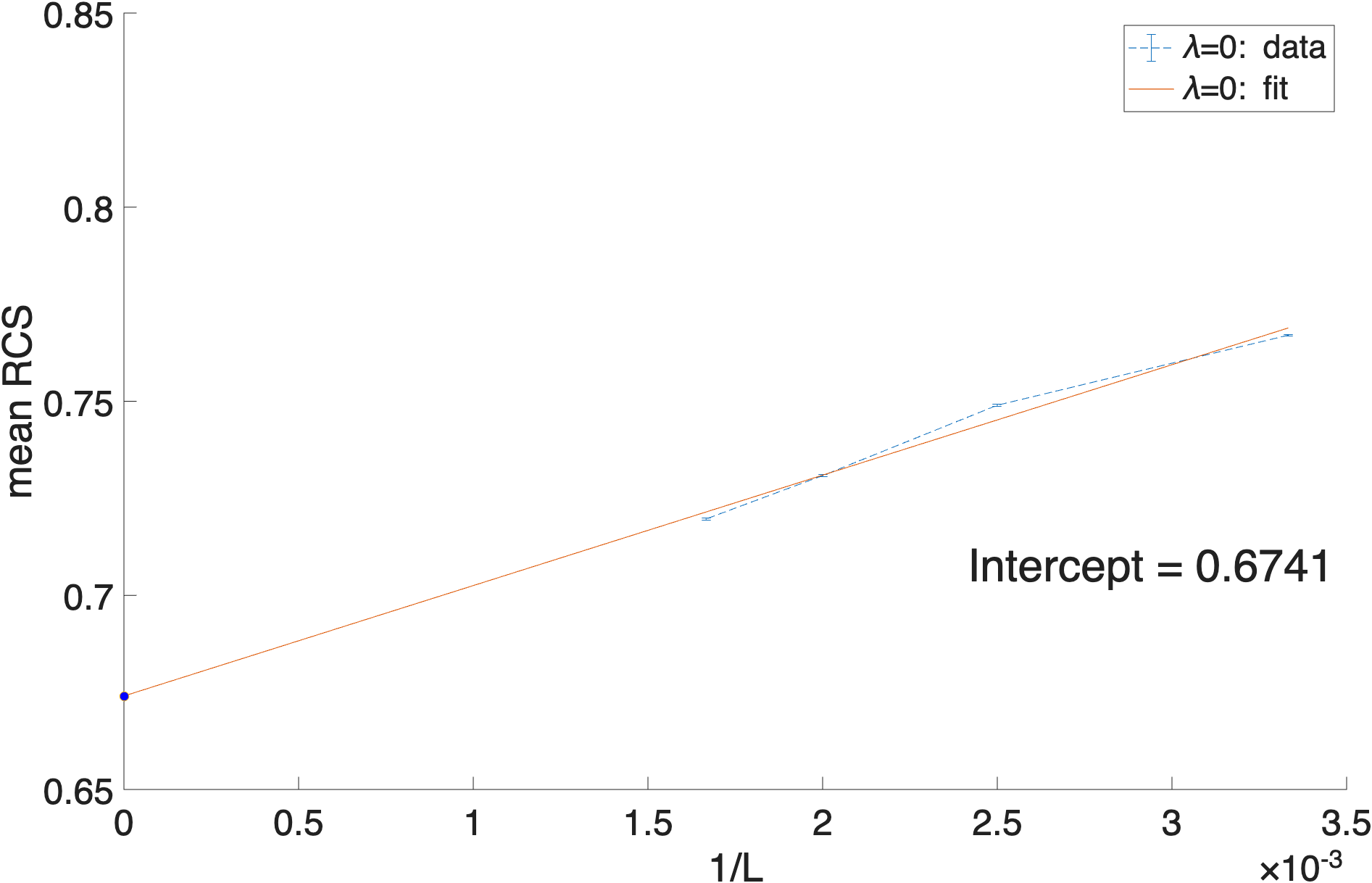}
        \label{fig:rcsn}%
    }
    \hfill
    \subfloat[]{%
        \includegraphics[width=0.9\linewidth]{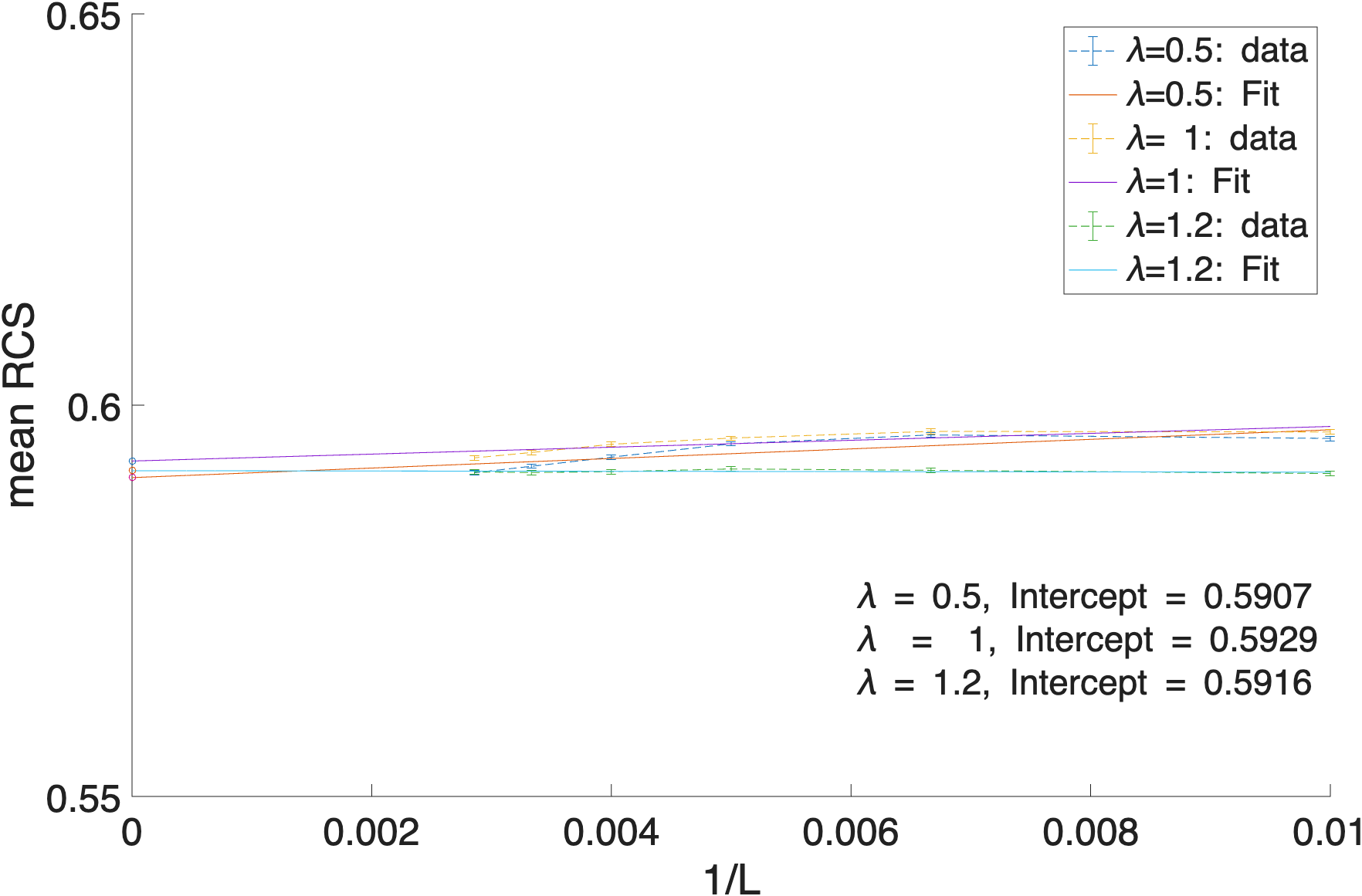}%
        \label{fig:rcst}%
    }

    \caption{\ref{fig:rcsn}: Mean RCS for single upright Dirac fermion, \ref{fig:rcst}: Mean RCS for single tilted Dirac fermion for $\lambda=0.5, 1, 1.2$. We have done linear fit to our data and extrapolate to extrapolate it to find the mean RCS for infinite system. With this we found $0.6741, 0.5907, 0.5929, 0.5916$ respectively for single node $\lambda=0, 0.5, 1, 1.2$ which are close to $0.6742$ and $0.5997$, mean  RCS for GSE and GUE, respectively.  }
\end{figure} 

 We have also analyzed another robust indicator of ergodic behavior, the RCS: $r_n = \frac{\min(\delta_n, \delta_{n+1})}{\max(\delta_n,\delta_{n+1})}$, where $\delta_n = E_{n+1} - E_n$. Its strength lies in being both local and scale-invariant, qualities that make it uniquely resistant to nonuniversal spectral variations. Unlike the conventional level spacing $s_n = E_n - E_{n+1}$, which measures the absolute distance between two neighboring energy levels, $r_n$ compares two consecutive spacings, $E_n$ and $E_{n+1}$, and highlights how they fluctuate relative to one another. In physical terms, this tells us whether neighboring energy levels tend to repel (as in chaotic or ergodic systems) or cluster (as in localized or integrable systems). Because the ratio only involves adjacent gaps, it depends purely on local correlations in the spectrum. For single Dirac fermion, as shown in Fig. \ref{fig:rcst}, we find the mean RCS to be around $0.59$, in close agreement with the theoretically predicted value of $0.5997$ for GUE.

\subsection{Two nodes}

Having examined the single-node system and identified several intriguing features in its transport behavior, we now turn to the case of two Dirac nodes. Unlike a single Dirac node, which is a topological object that can only arise on the surface of a three dimensional topological insulator in the presence of $\mathcal{T}$ symmetry, or at a quantum Hall plateau or Chern insulator transitions when $\mathcal{T}$ is broken, two Dirac nodes can naturally occur in ordinary 2D materials without any topological protection i.e., Borophene~\citep{islam2017signature, PhysRevB.94.165403, verma2017effect}. Moreover, 3D Dirac semimetals generically host pairs of tilted nodes, implying that their thin-film ~\citep{rizza2022extreme, PhysRevX.6.031021,chang2016prediction} realizations will exhibit effective 2D tilted Dirac fermions. In what follows, we explore how much of the delocalization physics observed in the single node model survives in the presence of two coupled nodes, and how inter-node scattering and symmetry considerations modify the overall transport behavior.

\begin{figure}
  \centering

    \subfloat[]{%
        \includegraphics[width=0.31\linewidth]{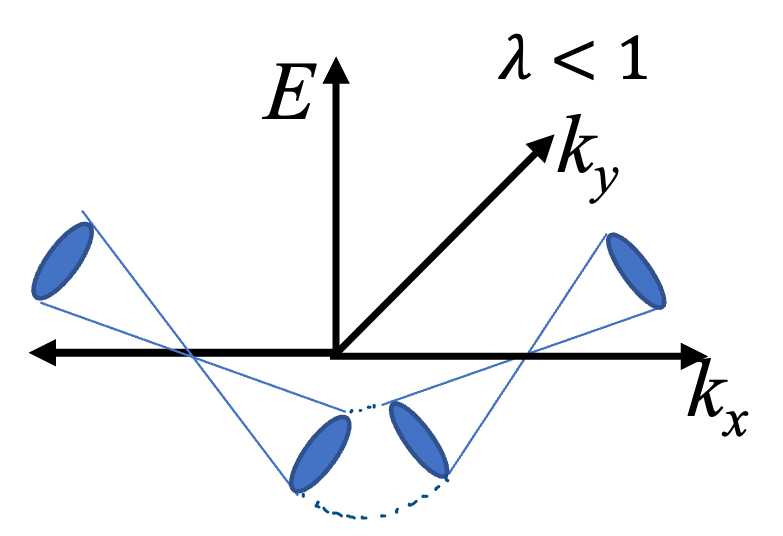}
        \label{fig:doublet1}%
    }
    \hfill
    \subfloat[]{%
        \includegraphics[width=0.31\linewidth]{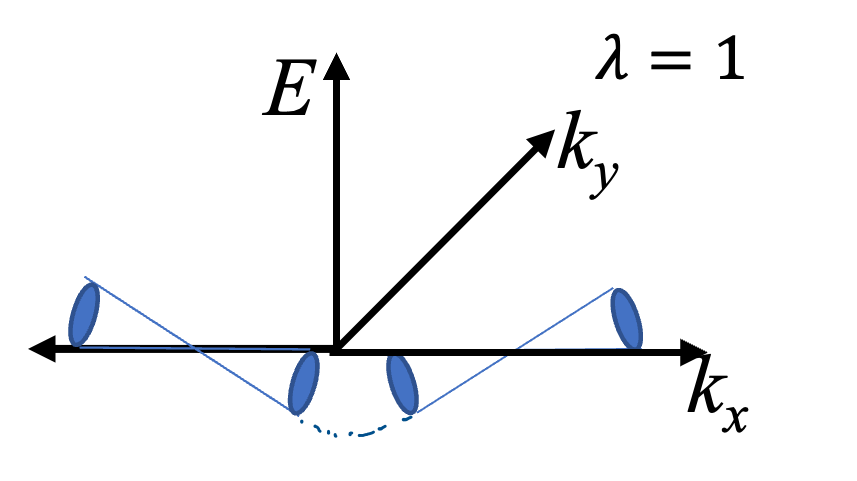}%
        \label{fig:doubletc}}%
    \hfill
    \subfloat[]{%
        \includegraphics[width=0.34\linewidth]{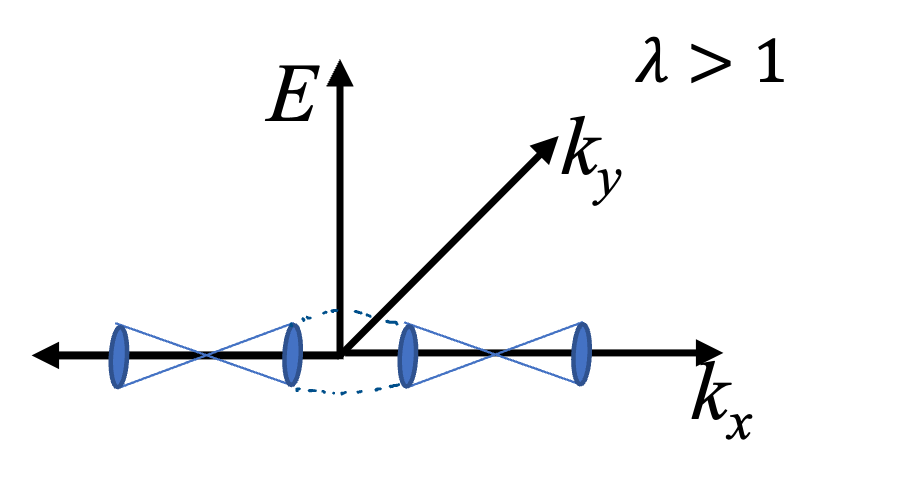}%
        \label{fig:doublet2}%
    }

\caption{ \ref{fig:doublet1}, \ref{fig:doubletc}, \ref{fig:doublet2} are showing type I, critical, and type II  Dirac node pairs, respectively. }
\label{fig:double}
\end{figure}

\subsubsection{Conductivity}
\begin{figure}
\centering
\subfloat[]{\includegraphics[width=0.9\linewidth]{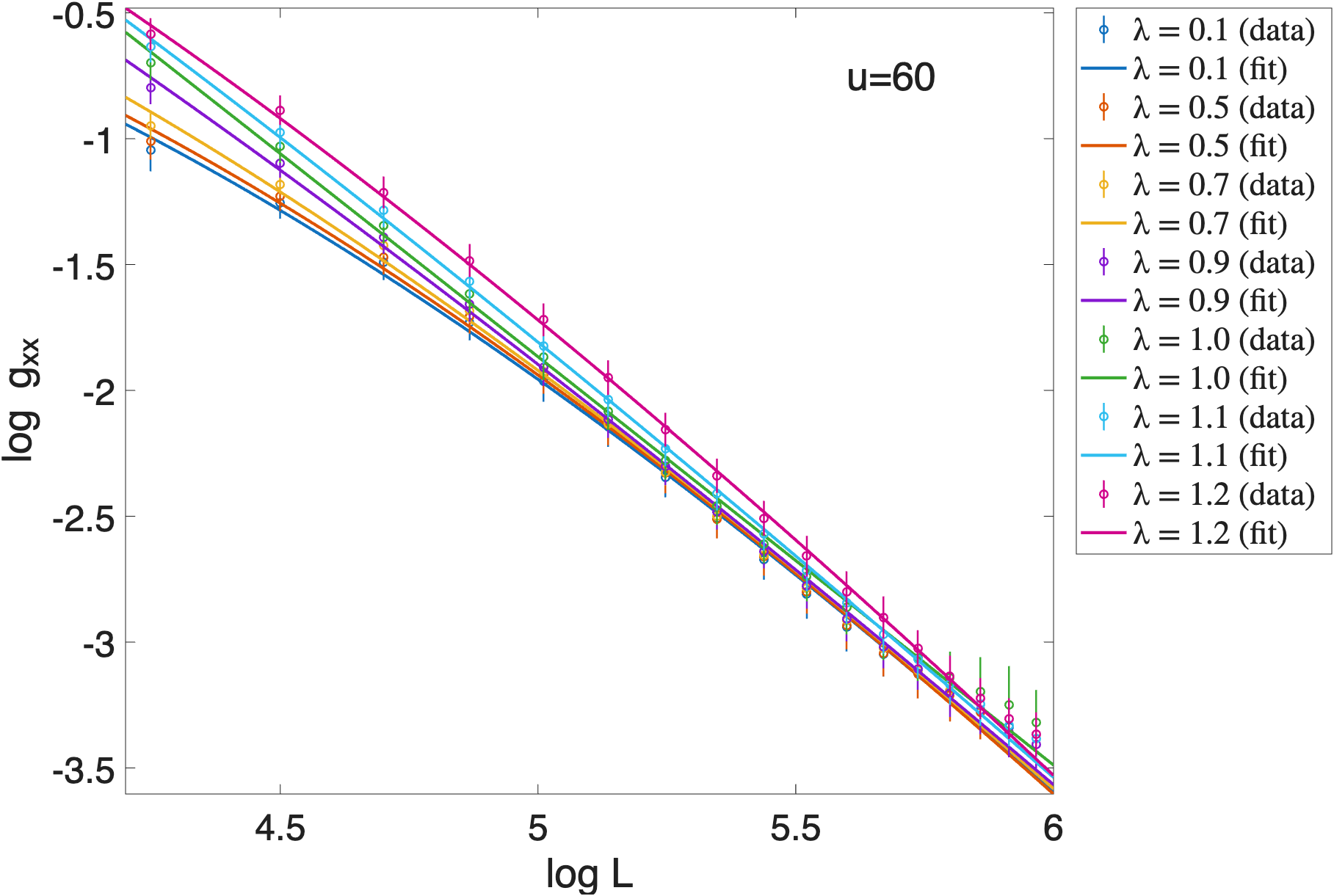}\label{fig:g60logxx}}

\subfloat[]{\includegraphics[width=0.9\linewidth]{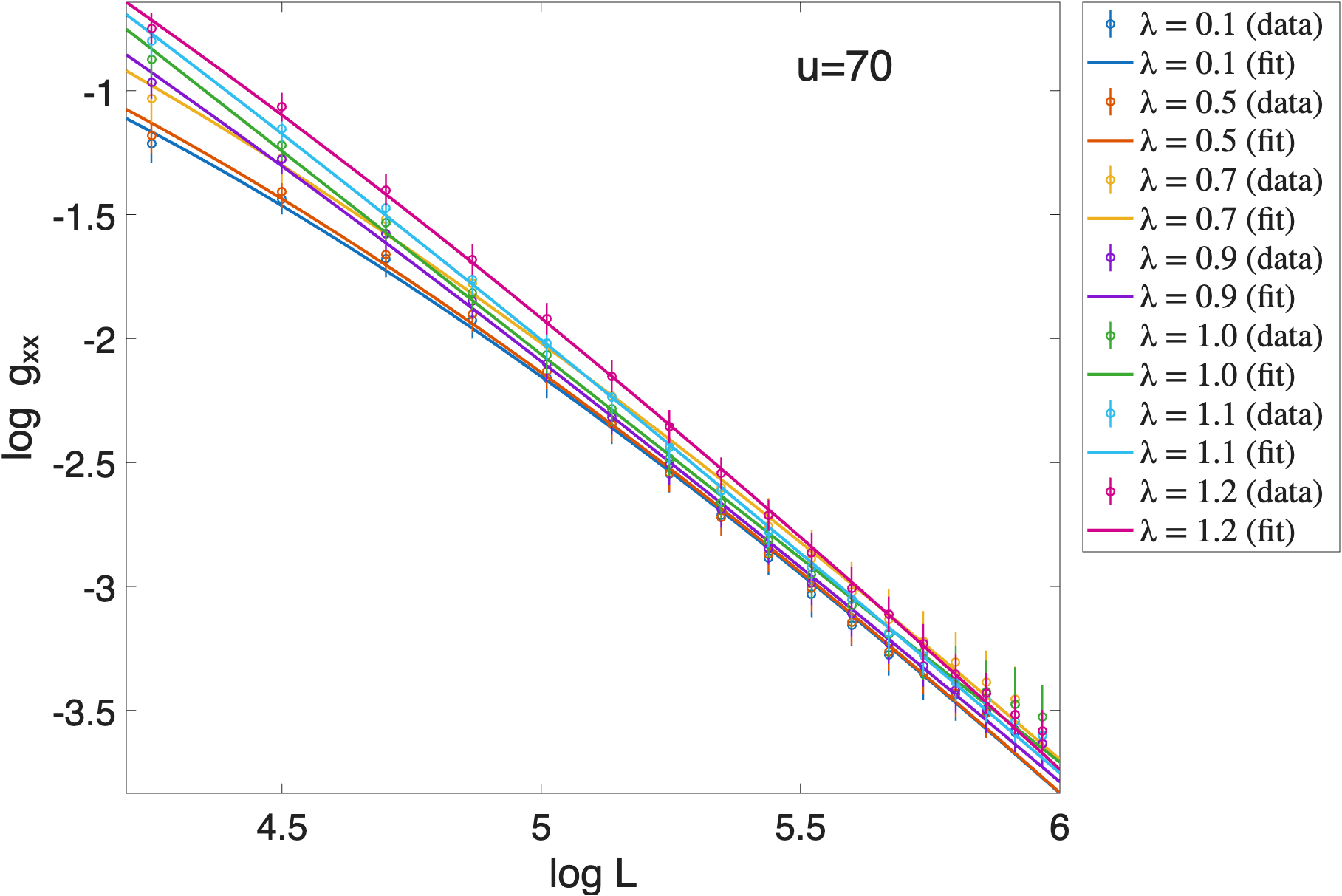}\label{fig:g70logxx}}

\subfloat[]{\includegraphics[width=0.9\linewidth]{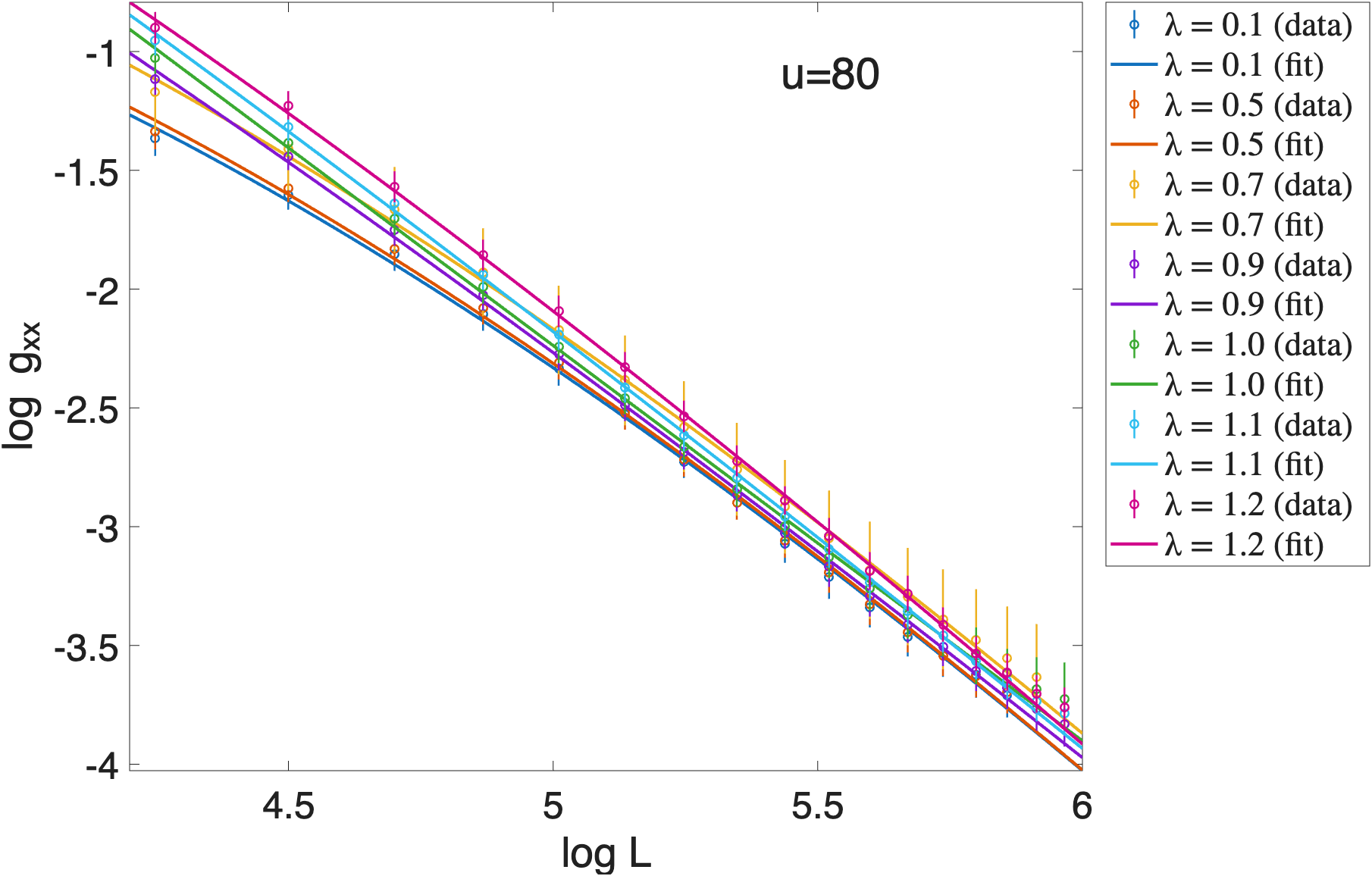}\label{fig:g80logxx}}
\caption{Fit of $g_{xx}$ of two Dirac nodes in $\log$-$\log$ scale. For each $\lambda$, we have fitted $ g_{xx} = e^{a_0} L^{a_1}(\log L)^{a_2} $. Fitting parameters $a_0, a_1, a_2$ has been shown in figure \ref{fig:a2xx}. }
\label{fig:g2xxfit}
\end{figure}

\begin{figure}
\centering
\subfloat[]{\includegraphics[width=0.9\linewidth]{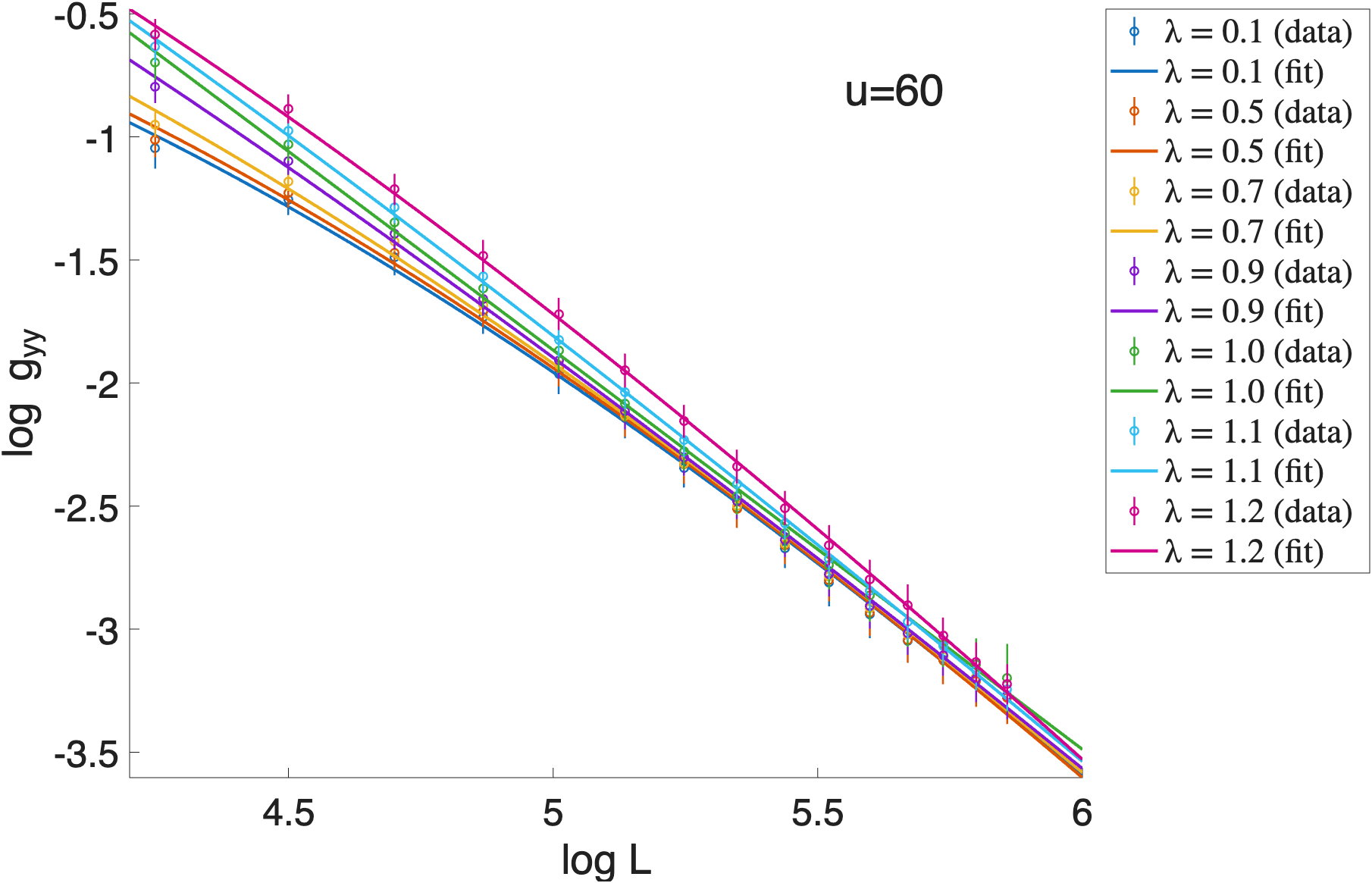}\label{fig:g60logyy}}

\subfloat[]{\includegraphics[width=0.9\linewidth]{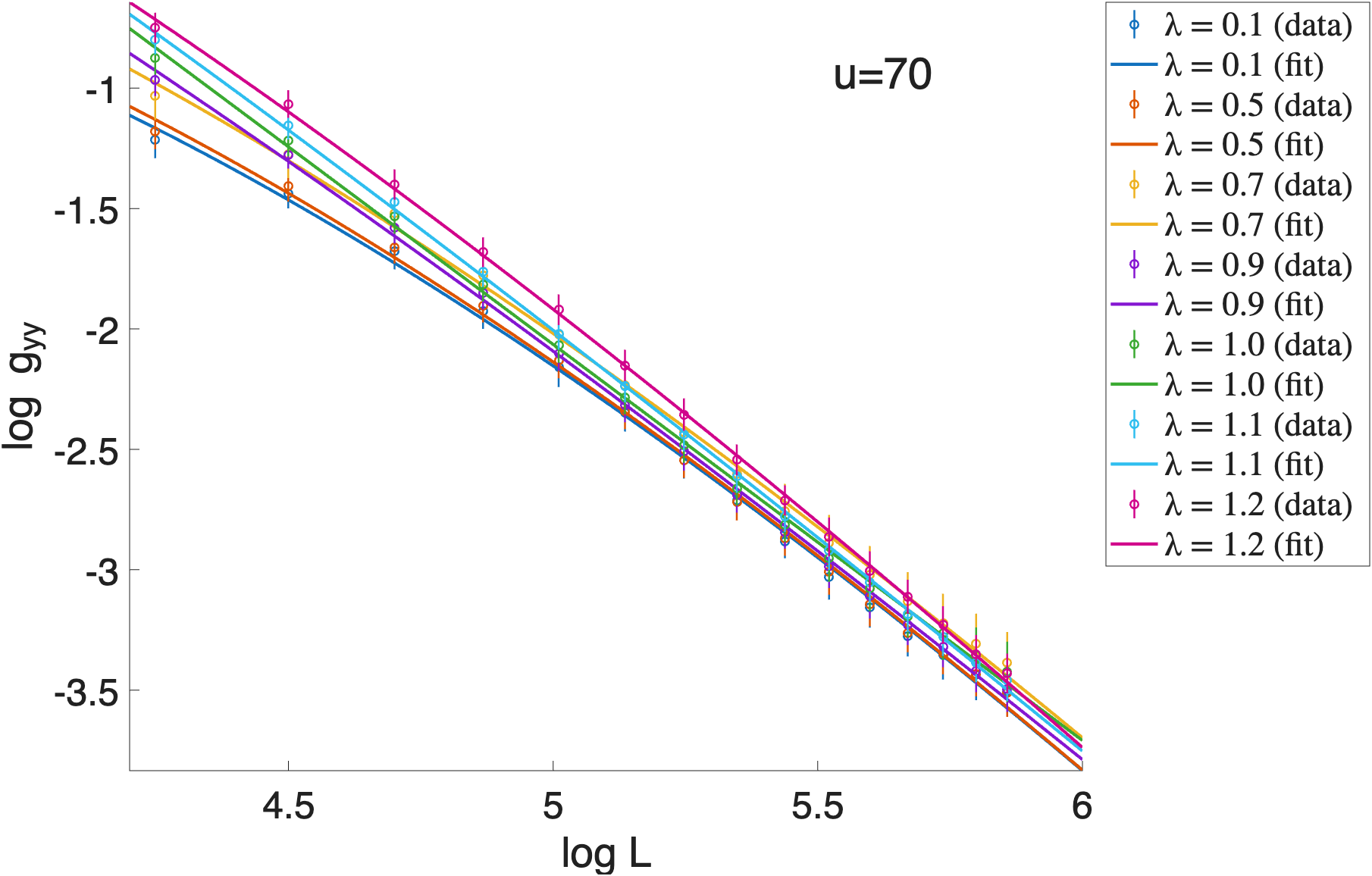}\label{fig:g70logyy}}

\subfloat[]{\includegraphics[width=0.9\linewidth]{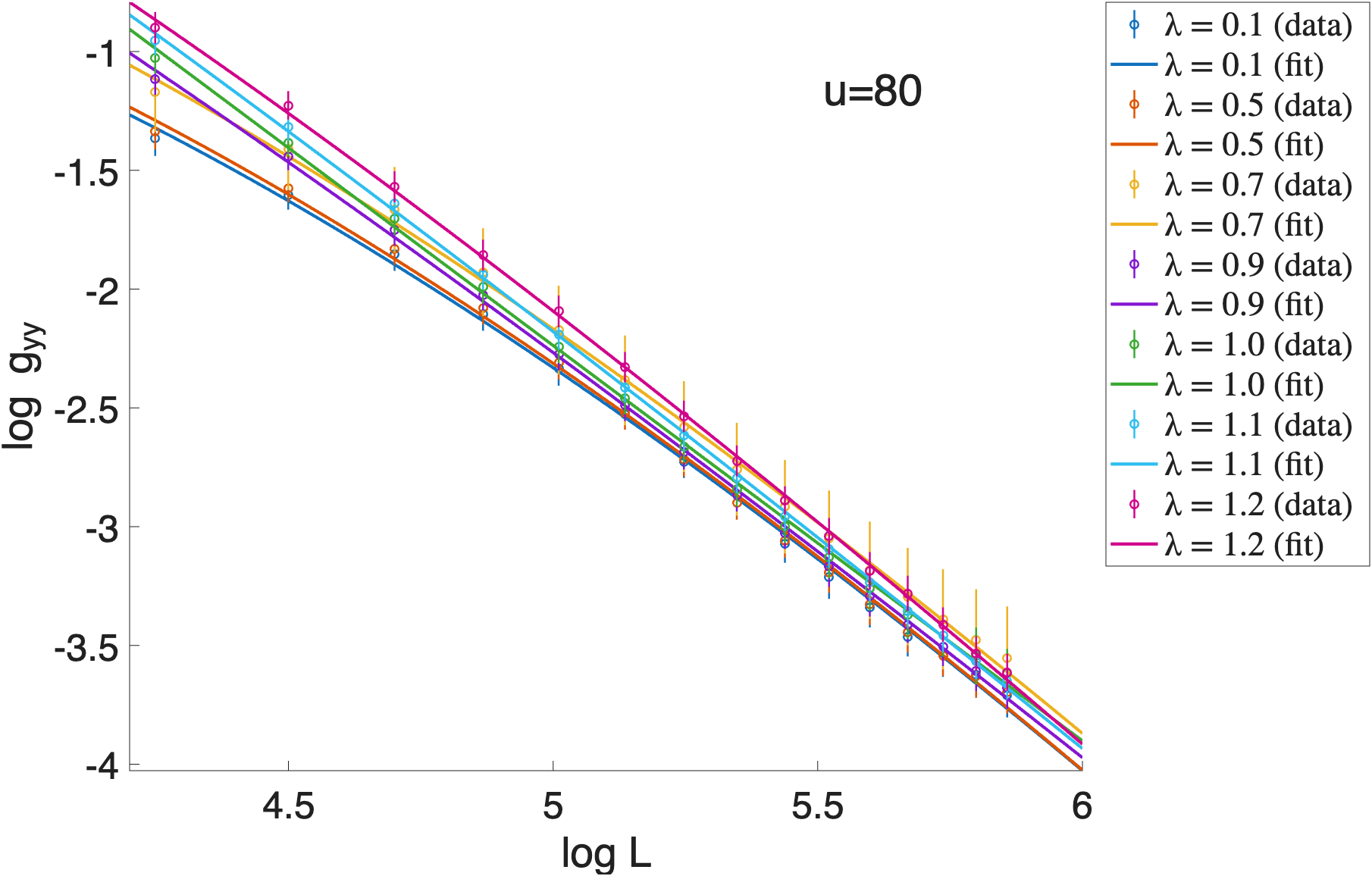}\label{fig:g80logyy}}
\caption{Fit of $\log g_{yy}$ of two Dirac nodes with $\log L$. For each $\lambda$, we have fitted $ g_{yy} = e^{a_0} L^{a_1}(\log L)^{a_2} $ with our data. Fitting parameters $a_0, a_1, a_2$ has been shown in Fig. \ref{fig:a2yy}. }
\label{fig:g2yyfit}
\end{figure}

\begin{figure}
 \centering
\subfloat[]{\includegraphics[width=0.9\linewidth]{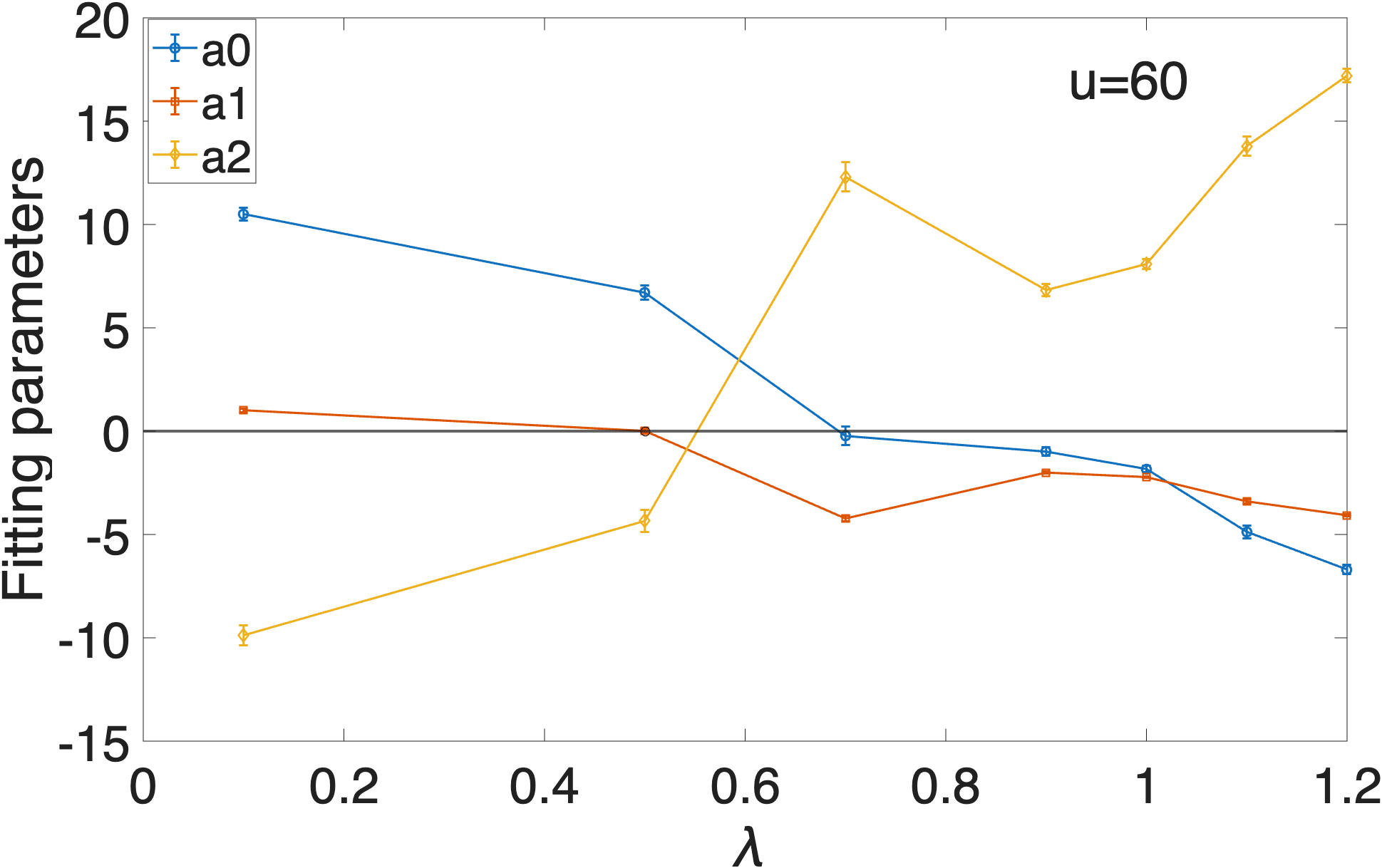}\label{fig:a2xx60}}

\subfloat[]{\includegraphics[width=0.9\linewidth]{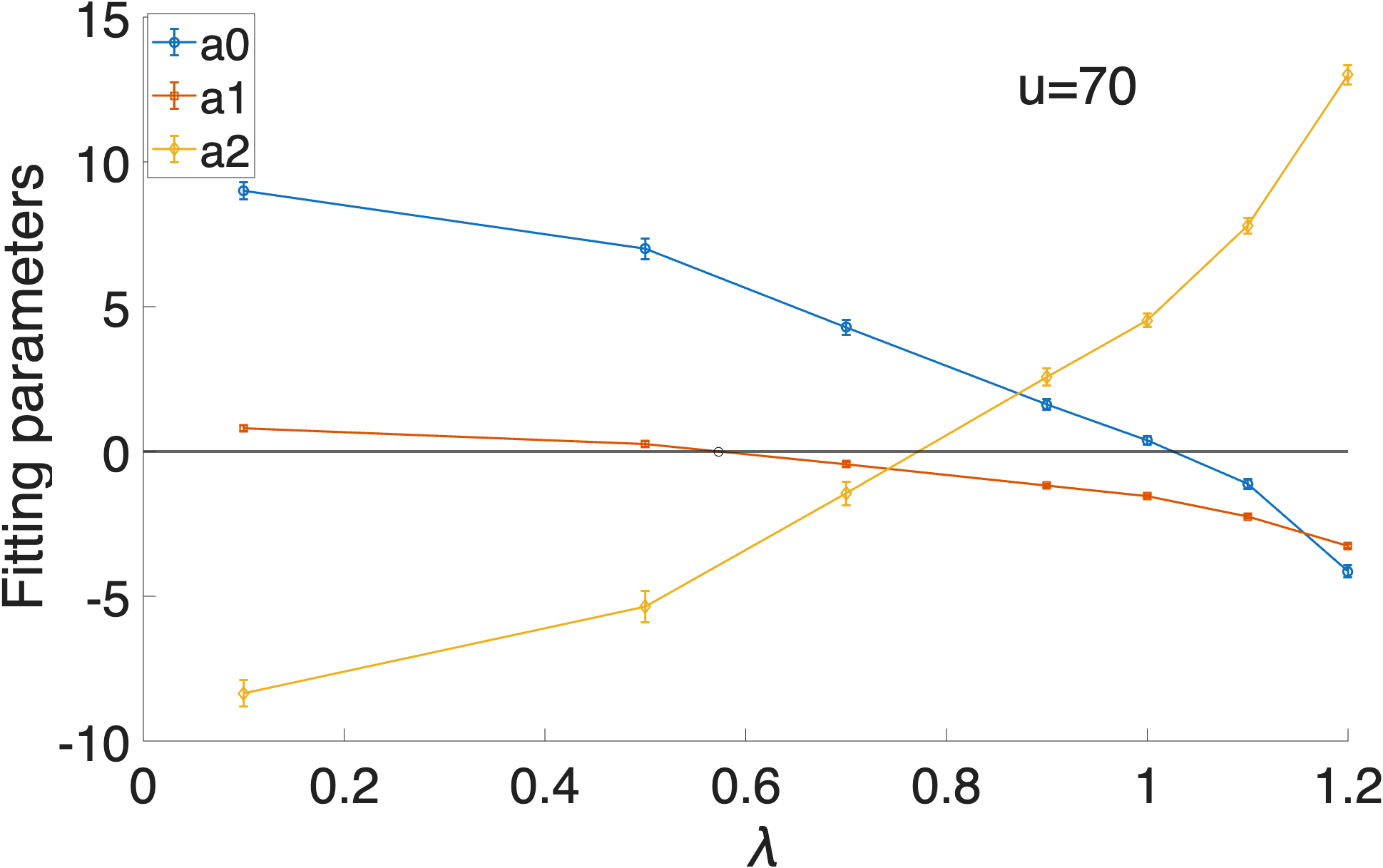}\label{fig:a2xx70}}

\subfloat[]{\includegraphics[width=0.9\linewidth]{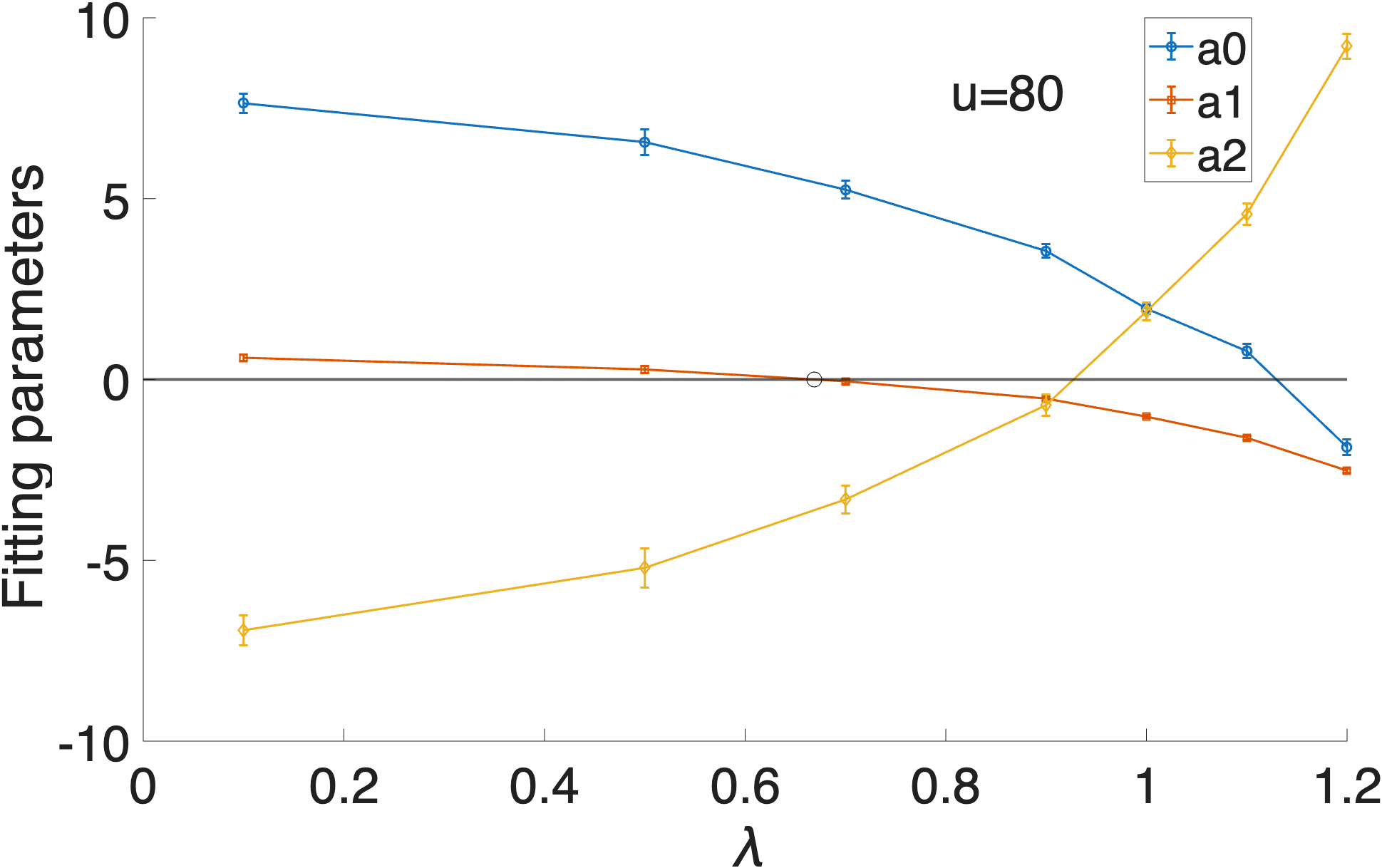}\label{fig:a2xx80}}   

\caption{Fitted parameters $a_0, a_1, a_2$ for $g_{xx}$ from the plots of Fig. \ref{fig:g2xxfit}. Though all three parameters are changing sign, the most important feature to note is the sign change of $a_1$ that determines the conductivity at thermodynamic limit.  }
\label{fig:a2xx}
\end{figure}

\begin{figure}
    
\subfloat[]{\includegraphics[width=0.9\linewidth]{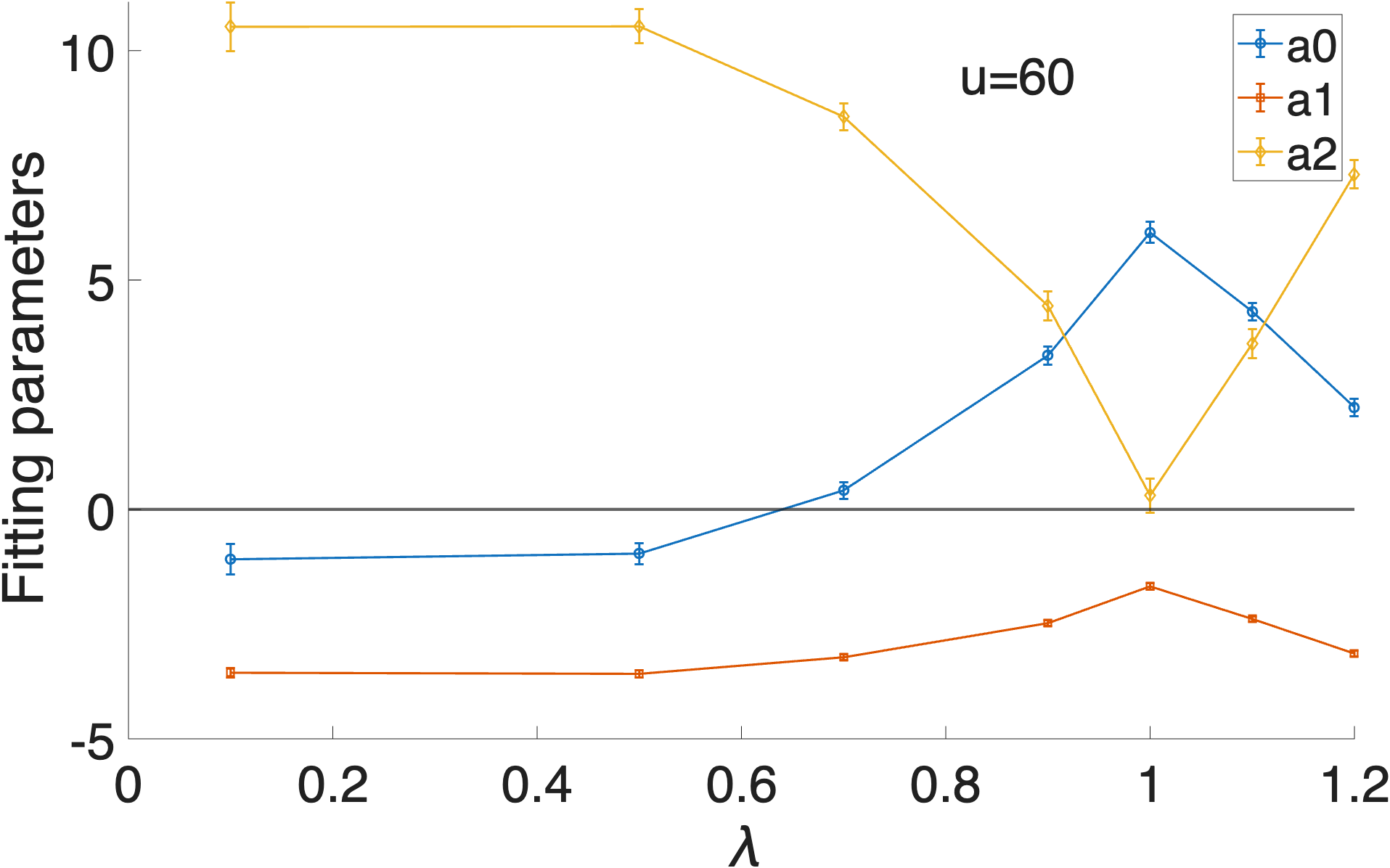}\label{fig:a2yy60}}

\subfloat[]{\includegraphics[width=0.9\linewidth]{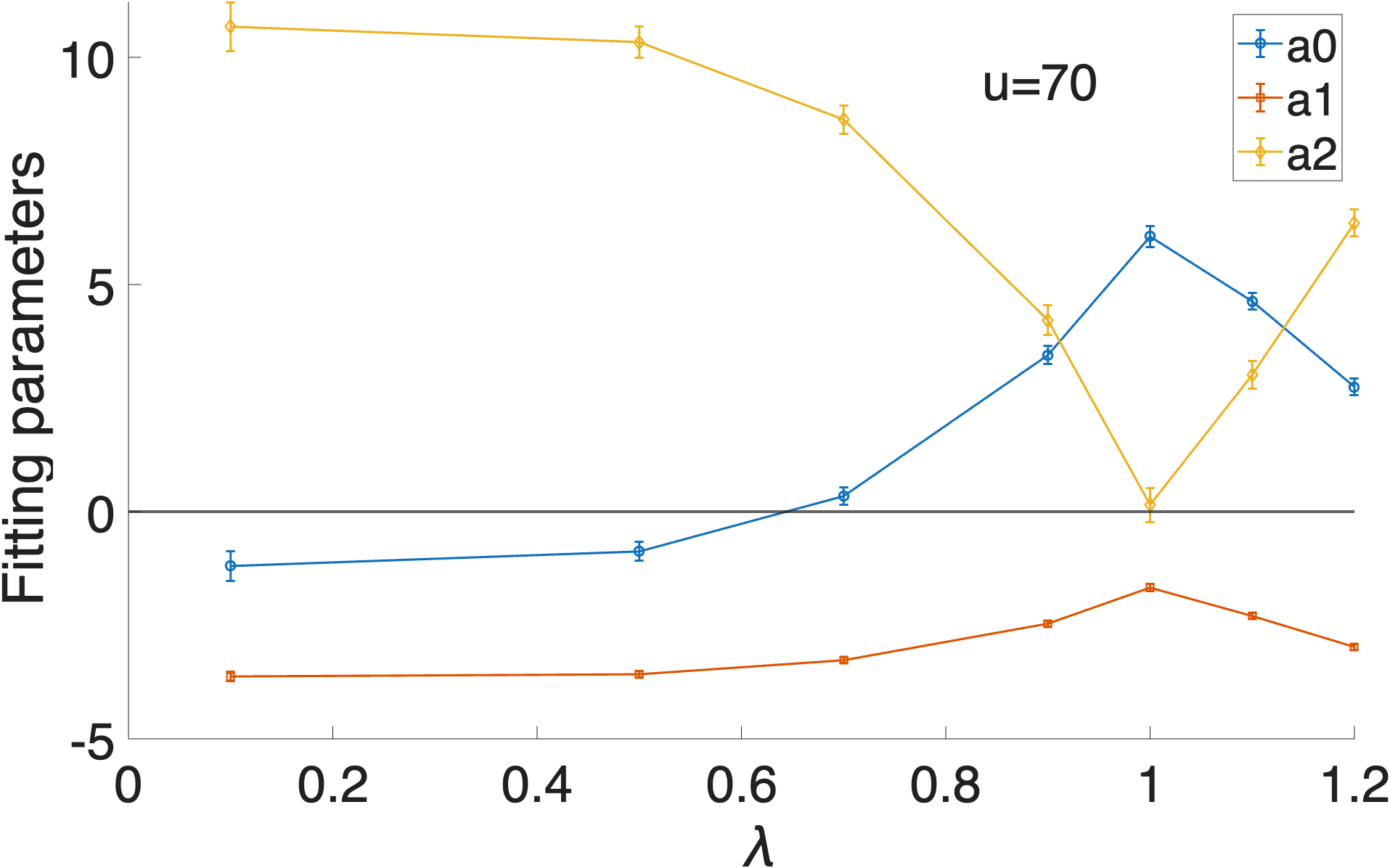}\label{fig:a2yy70}}

\subfloat[]{\includegraphics[width=0.9\linewidth]{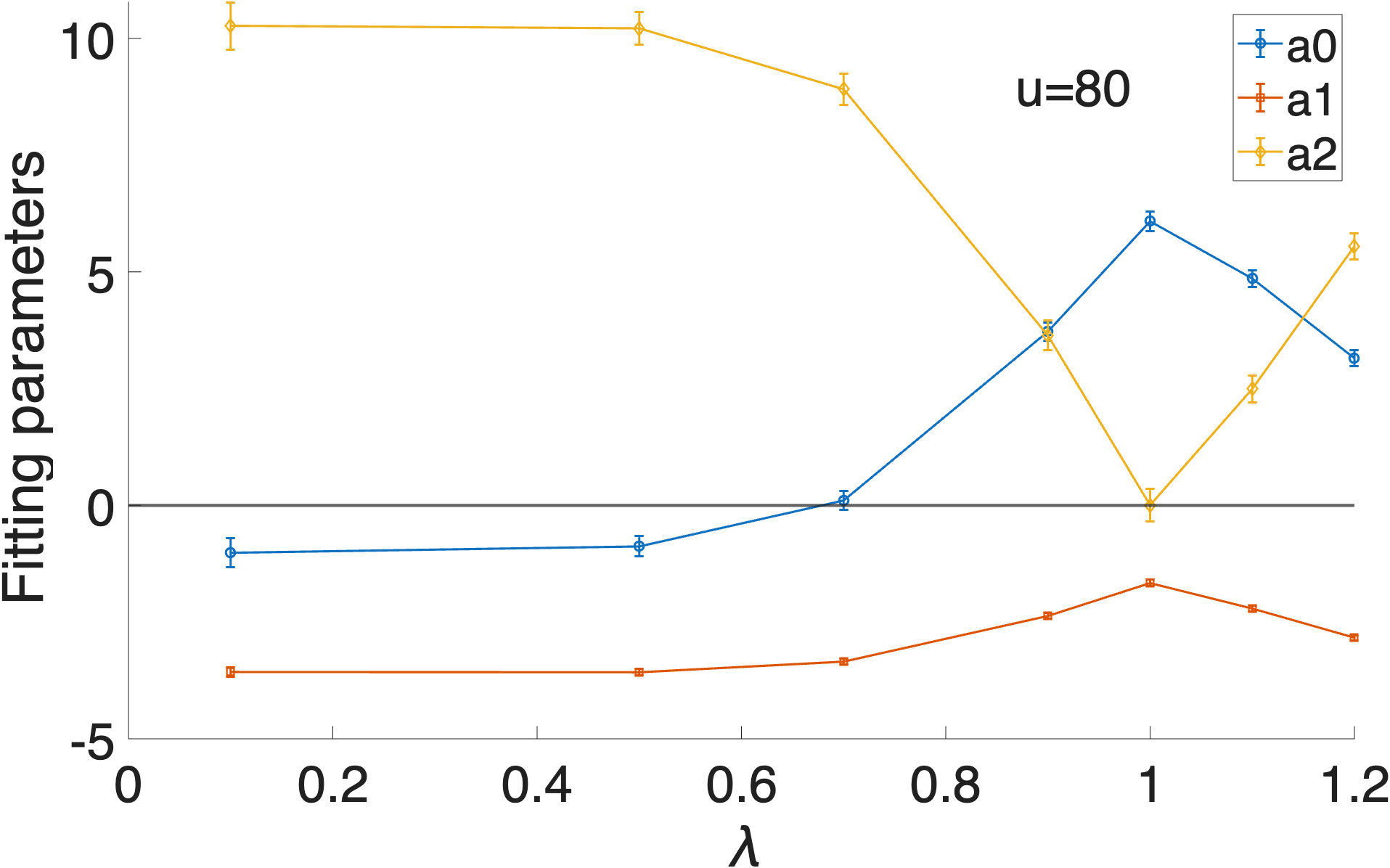}\label{fig:a2yy80}}
\caption{Fitted parameters $a_0, a_1, a_2$ for $g_{yy}$ from the plots of Fig. \ref{fig:g2yyfit}. Unlike the case for $g_{xx}$, here $a_1$ does not change sign. Instead, interesting behavior of the parameters is shown at $\lambda=1$. More precisely, $a_1$ is maximum at $\lambda=1$, which implies $\beta$ function is maximum for $\lambda=1$ in the thermodynamic limit. }
\label{fig:a2yy}
\end{figure}

As illustrated in Fig.~\ref{fig:g2xxfit} and Fig.~\ref{fig:g2yyfit}, the data for $\log{g_{ii}}$ are well described by the fitting form
\begin{equation}
 g_{ii} = e^{a_0} L^{a_1}(\log L)^{a_2} 
\end{equation}
that captures both the leading power-law scaling and the subleading logarithmic corrections expected in two-dimensional systems with marginal disorder. Equivalently, \begin{equation}\beta =  a_1 + \frac{a_2}{\log L}\approx a_1\left(1+\frac{a_2}{\log g}\right)\end{equation}
for $|\log g|\gg1$. The extracted coefficients $a_0, a_1, a_2$ for different values of tilt and disorder strength are presented in Fig.~\ref{fig:a2xx} and Fig.~\ref{fig:a2yy}. These coefficients provide valuable insight into the asymptotic scaling behavior of the conductivity.
In the thermodynamic limit $L\to\infty$, $\beta$ approaches the constant value $a_1$. Thus, the sign of $a_1$ directly reflects the asymptotic scaling behavior of $g_{ii}$: for $a_1>0$, the conductivity increases with system size, indicating extended or delocalized states, while for $a_1 <0$, the conductivity decreases, signifying localization.
Fig.~\ref{fig:a2xx} describes the behavior of $g_{xx}$, clearly showing a sign change in $a_1$ as the tilt parameter is increased, implying that the $\beta$ function also switches from positive to negative. This sign reversal of $a_1$ therefore marks a fundamental change in the scaling flow of the conductivity: from a regime where the system flows toward higher conductivity with increasing size (metallic/delocalized phase) to one where it flows toward smaller conductivity (insulating/localized phase). The point where $a_1=0$ corresponds to a scale-invariant state, identifying the critical tilt that separates the delocalized and localized regimes. This behavior suggests a the possibility of a tilt-driven localization–delocalization crossover -- or possibly a transition -- in the two-node Dirac model along the tilt direction.


In contrast to $g_{xx}$, the scaling behavior of $g_{yy}$ does not exhibit a clear sign reversal in the coefficient $a_1$, and thus no indication of a localization–delocalization transition or crossover is observed from its asymptotic flow. Nevertheless, the extracted fitting parameters display a distinct and systematic dependence on the tilt parameter. As shown in Fig.~\ref{fig:a2yy}, all three coefficients $ a_0, a_1, a_2$ exhibit pronounced extremal behavior near the critical tilt 
$\lambda=1$. In particular, $a_1$ is maximum at $\lambda=1$, and this implies that $\beta$ function is maximum at $\lambda = 1$. This suggests that even though $g_{yy}$ does not undergo a qualitative change in scaling behavior, its amplitude and subleading corrections are strongly affected by the tilt-induced anisotropy of the band structure, especially near the boundary of type I and type II Dirac nodes.

 In the single node model, internode scattering is absent by construction; with only one Dirac point, there is no momentum transfer that can scatter states between distinct nodes. This assumption is physically sensible for an upright or weakly tilted Dirac cone, where the two valleys of the full lattice remain well separated in momentum space and internode processes are strongly suppressed. Near the critical tilt, however, this picture becomes increasingly fragile. Strong tilting of the cone reduces the effective momentum separation between nodes, opening scattering channels with essentially zero momentum transfer. In this regime, a realistic description must treat both nodes explicitly, which is precisely what the two-node model provides. Once the nodes are coupled, the topological protection associated with a single isolated Dirac point is lost, enabling constructive interference of backscattered paths and ultimately driving the system toward localization.

\subsubsection{Level Statistics }

\begin{figure}
    \centering
    \includegraphics[width=0.9\linewidth]{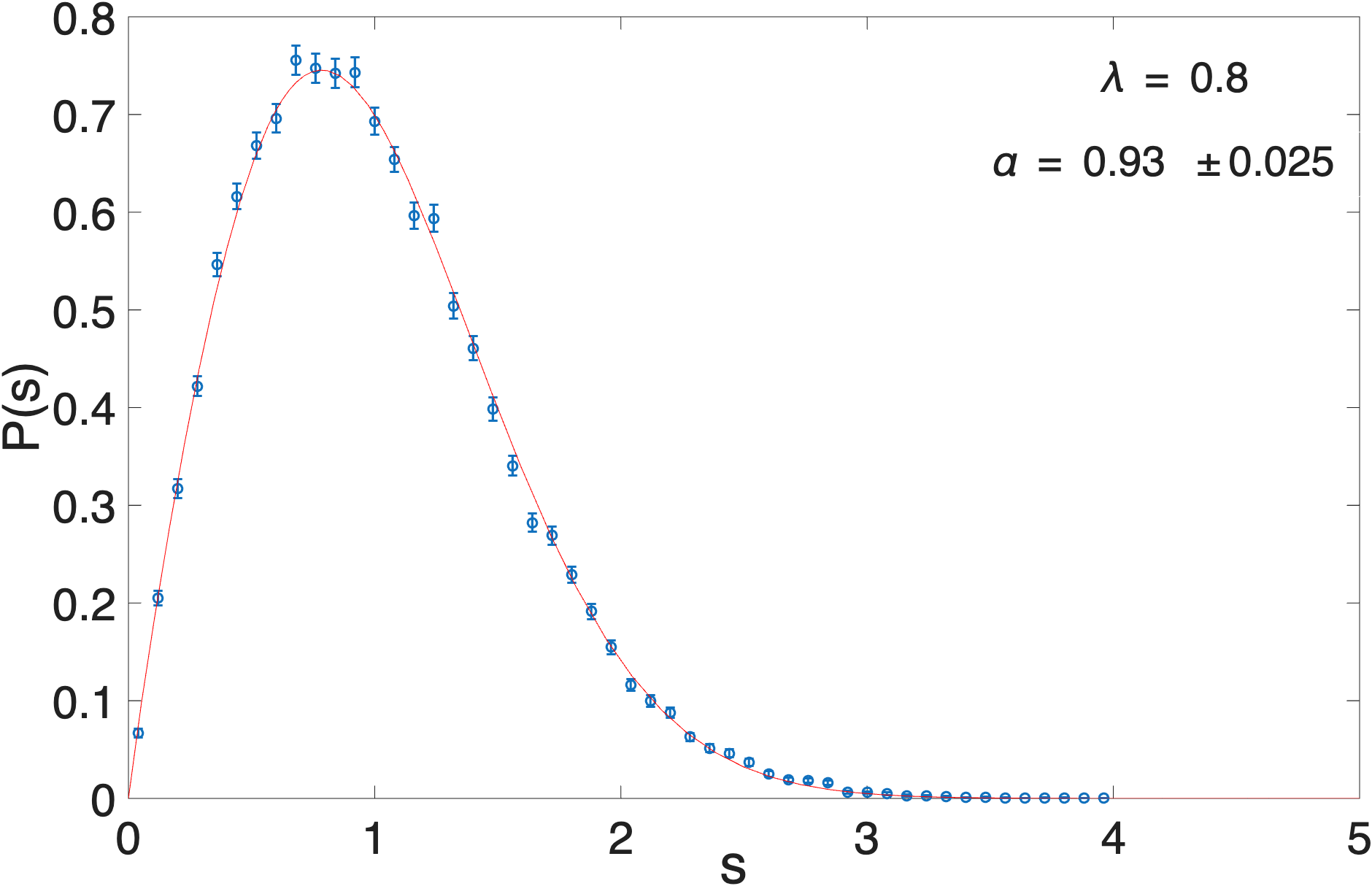}
    \caption{Level statistics for tilted two-node Dirac system. Fitted value for the exponent $\alpha\sim 0.93$, in close agreement with theoretical prediction of GOE from symmetry considerations.}
    \label{fig:ls2}
\end{figure}

\begin{figure}
    \centering
    \includegraphics[width=0.9\linewidth]{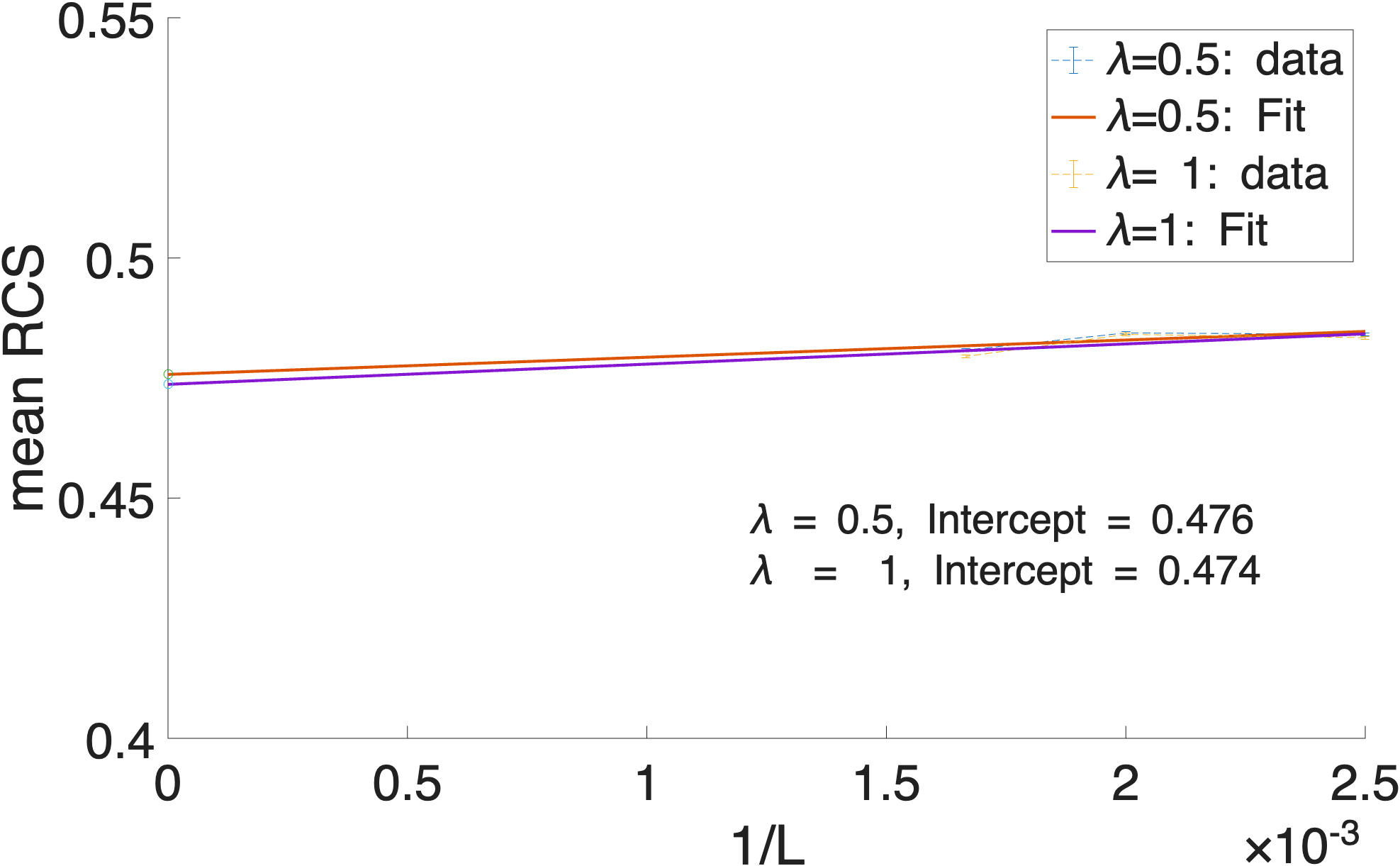}
    \caption{Mean RCS for tilted two-node Dirac system. We have followed similar method as single node, and got mean RCS to be around 0.47 for $L\to\infty$.}
    \label{fig:rcs2}
\end{figure}

In the two-node model, we deployed similar methods as we did for a single Dirac fermion. In this case, as shown in Fig. \ref{fig:ls2}, we got $\alpha\sim0.93$ which matches closely with GOE ($\alpha=1$). For RCS, we got the mean RCS to be around $0.45$, as shown in Fig. \ref{fig:rcs2}. This shows some deviation from the theoretically predicted value of $0.536$. 

\section{Conclusion}
In summary, we have investigated the combined effects of disorder strength and band tilt on 2D disordered Dirac systems. Our results show that both spectral and transport characteristics remain largely insensitive to the disorder amplitude $u$ for systems with either a single or double Dirac node. For a single Dirac node, the conductivity along the tilt direction exhibits a pronounced enhancement at the critical tilt corresponding to the type-I to type-II transition, followed by a gradual suppression with further tilting. The conductivity in direction orthogonal to the tilt, in contrast, increases monotonically with tilt, and the system remains delocalized across all tilt values. Spectral analysis of a single tilted Dirac fermion, based on level statistics and mean RCS conforms to GUE distribution.

For systems with two Dirac nodes, however, the behavior differs qualitatively. While finite-size calculations indicate localization, the $\beta$-function in the thermodynamic limit changes sign with tilt for transport along the tilt direction, suggesting a tilt-driven delocalization–localization transition or crossover. In contrast, localization persists in the orthogonal direction, even in the thermodynamic limit. Spectral correlation, i.e., level statistics and RCS, however, reveals delocalization (GOE). 

\acknowledgments
We are grateful to Matthew S. Foster and Kentaro Nomura for insightful discussions. This work was supported by the Department of Energy grant number DE-SC0022264.

\newpage
\appendix
\onecolumngrid
\section{Density of states for clean Dirac fermion}
\label{app:dos}
Here we will derive the DOS for a tilted, clean Dirac fermion.
Assuming tilt in the $x$ direction, our Hamiltonian for a single Dirac node reads 
\begin{equation}
    H(\boldsymbol{k}) = \lambda k_x + k_x\sigma_x + k_y\sigma_y
\end{equation}
with $v_F = 1$, and $\lambda$ dictates tilting of the node. Dispersion is given by $E_{\pm} = \lambda k_x \pm \sqrt{k_x^2 + k_y^2}$. The density of states is given by 
\begin{equation}
    \rho(E) = \int \frac{d^2k}{(2\pi)^2}(\delta(E-E_+) + \delta(E-E_-))
    \label{eq:dos2}
\end{equation}
Going to polar coordinate $k_x = k\cos{\theta}, k_y = k\sin{\theta}$, $d^2k = kdkd\theta$. Now, $\delta(E - E_{\pm}) = \delta(E - k(\lambda\cos{\theta} \pm 1))$. Now solving the $k$ integral using the delta function, 
\begin{equation}
    \rho(E) = \sum_{s=\pm}\frac{|E|}{2\pi}\int_0^{2\pi}\frac{d\theta}{(2\pi)^2}\left|\frac{E}{s + \lambda\cos{\theta}}\right|^2 \label{eq:dos_gen}
\end{equation}
Solving these, for type I Dirac node, we get 
\begin{equation}
    \rho(E) = \frac{2|E|}{\pi (1-\lambda^2)^{3/2}}
\end{equation}
But in the type II region, since $\lambda > 1$, Eq. (\ref{eq:dos_gen}) will have divergences for certain values of $\theta$, and to obtain a closed form for the DOS we will use a momentum cut off $\Lambda$. We will begin with Eq. (\ref{eq:dos}),
\begin{equation}
    \rho(E,\Lambda)=\frac{1}{(2\pi)^2}\sum_{s=\pm}\int_0^{2\pi}d\theta\int_0^{\Lambda}dkk\delta(E-k(s+\lambda\cos\theta))
\end{equation}
Evaluating the $k$ integral using the delta function produces
\begin{equation}
    \rho(E,\Lambda)=\frac{|E|}{(2\pi)^2}\sum_{s=\pm}\int_0^{2\pi}\frac{d\theta}{A_s(\theta)^2}\Theta(A_s(\theta))\Theta(\Lambda A_s(\theta)-E)
\end{equation}
where $A_s(\theta)=s+\lambda\cos\theta$.
These two $\Theta$- functions are coming from the facts that $k>0$ and $k<\Lambda$, and will restrict the limit of the $\theta$ integral. Using these two $\Theta$ functions, we found the range of $\theta$: $\cos\theta>\frac{E/\Lambda-s}{\lambda}$. Assume $c_s(\theta)=\frac{E/\Lambda-s}{\lambda}$, so we have $\theta_s(E)=\cos^{-1}(c_s(E))$. Now our range of the integral is over $\theta\in[0,\theta_s]\cup[2\pi-\theta_s,2\pi]$. With this we can write
\begin{equation}
    \rho(E,\Lambda)=\frac{E}{2\pi^2}\left[I_{+}(E) + I_{-}(E)\Theta(\Lambda(\lambda-1)-E)\right]
\end{equation}
where, $J_s(E)=\int_0^{\theta_s(E)}\frac{d\theta}{(\lambda+s\cos\theta)^2}$. Now the task is to compute $J_s(E)$.
By using integration by parts and some manipulation, we can write
\begin{equation}
    \int\frac{d\theta}{(a+b\cos\theta)^2}=\frac{\sin\theta}{(a^2-b^2)(a+b\cos\theta)} + \frac{a}{a^2-b^2}\int\frac{d\theta}{a+b\cos\theta}
\end{equation}
and 
\begin{equation}
    \int\frac{d\theta}{a+b\cos\theta}=\frac{2}{\sqrt{b^2-a^2}}\tanh^{-1}\left(\sqrt{\frac{b-a}{b+a}}\tan{\frac{\theta}{2}}\right), |b|>|a|
\end{equation}
With these two, we can write $I_s(E)$ explicitly
\begin{equation}
    I_s(E)=\frac{2\sin\theta_s}{(s^2-\lambda^2)(s+\lambda\cos\theta_s)} + \frac{2s}{(s^2-\lambda^2)\sqrt{\lambda^2-s^2}}\tanh^{-1}\left( \sqrt{\frac{\lambda+s}{\lambda-s}}\tan{\frac{\theta_s}{2}}\right)
\end{equation}
with $s=\pm $.

At $E=0$, we recover simple form for $\rho(0,\Lambda)$
\begin{equation}
    \rho(0,\Lambda) = \frac{\Lambda}{\pi^2(\lambda^2 -1)}
    \label{eq:dosE0type2}
\end{equation}
This shows that DOS of a clean Dirac fermion is vanishingly small at low energies, $E\to0 $, however it grows as the node is more and more tilted, and beyond critical tilt, i.e., for $\lambda>1$ DOS becomes finite even at $E=0$ as shown in Eq. (\ref{eq:dosE0type2}).
\section{Behavior of conductivity with tilt}
\label{app:analytics}
In this section, we provide perturbative analytical expressions that support some of the simulations performed in the main text. For that, we will begin this section with Eq.~\ref{eq: kubo}, and derive Eq. \ref{eq:kubotr} from it. Then from Eq. \ref{eq:kubotr}, we will show explicit tilt dependence of $g_{xx}$.

We will start writing the Kubo formula in Green's function form
\begin{equation}
    g_{xx}=-\frac{e^2\hbar}{2\pi L^2}\intop dE \frac{\partial f}{\partial E}\text{tr}\left[v_xG^{R}(E)\Gamma_xG^A(E)\right]
\end{equation}
where $G^{R/A}(E)=\frac{1}{E-H\pm\frac{i\hbar}{2\tau}}$ are disorder-averaged Green's function, $\Gamma_x$ is the dressed current vortex. In the eigenbasis $|n\rangle$ of $H$, $G^{R/A}_n(E)=\frac{1}{E-E_n\pm\frac{i\hbar}{2\tau_n}}$ where $\tau_n$ is single particle life-time $\frac{1}{\tau_n}=\frac{2\pi}{\hbar}\sum_{m}|\langle n|V|m\rangle|^2\delta(E_n-E_n)$. Now, in the weak scattering limit (where $n\sim m$), 
\begin{equation}
    \intop dE_n\left(-\frac{\partial f}{\partial E_n}\right)G^R_n(E)G^A_m(E)\sim-\frac{2\pi\tau_n}{\hbar}\frac{\partial f}{\partial E_n}\delta(E_n-E_m)
\end{equation}
With this, we get
\begin{equation}
    g_{xx}=-\frac{1}{L^2}\sum_n\frac{\partial f}{\partial E_n}\tau_n S_n
    \label{eq:kuboundressed}
\end{equation}
where $S_n=\sum_m|\langle n|v_x|m\rangle|^2\delta(E_n-E_m)$. In writing Eq. (\ref{eq:kuboundressed}), we have ignored vortex corrections by writing $v_x$. To incorporate this, we have to compute
\begin{equation}
    g_{xx}=-\frac{e^2\hbar}{2\pi L^2}\intop dE \frac{\partial f}{\partial E}\text{tr}\left[\Gamma_xG^{R}(E)\Gamma_xG^A(E)\right]
\end{equation}
where, $\Gamma_x$ is the dressed vertex, and can be written, in as
\begin{equation}
    \Gamma_x=v_x+\langle V G^R\Gamma_xG^AV\rangle
\end{equation}
where the $\langle..\rangle$ is disorder averaged.
As shown in Chapter 9 of \citep{bruus2004many}, we can replace $\Gamma_x$ by bare vertex $v_x$, $\Gamma_x=v_x\tau^{\text{tr}}/\tau$, where $\tau^{\text{tr}}$ is the transport relaxation time. With vertex corrections,
\begin{equation}
    S_n'=\sum_m|\langle n|\Gamma_x|m\rangle|^2\delta(E_n-E_m)
\end{equation}
Again in the weak disorder limit, we can approximate
\begin{equation}
    \langle n|\Gamma_x|m\rangle\sim\Gamma_x\delta_{n,m}=v_x^n\frac{\tau_n^{\text{tr}}}{\tau_n}\delta_{n,m}
\end{equation}
With this we can write
\begin{equation}
    \tau_nS'_n\sim (v_x^n)^2\tau_n^{\text{tr}}\rho(E_n)L^2
\end{equation}
Putting this into Eq. (\ref{eq:kuboundressed}), and converting sum to integral $\frac{1}{L^2}\sum_n(...)\to\intop dE\rho(E)\langle..\rangle_E$, we arrive at (up to some overall constants) Eq. (\ref{eq:kubotr}) 
\begin{equation}
    g_{xx}=\int dE_n\left(\frac{-\partial f}{\partial E_n}\right)\rho(E_n)\langle(v_x^n)^2\tau_n^{\text{tr}}\rangle_{E_n}
\end{equation}
Now, we can write $\rho(E)\langle A\rangle_E=\sum_nA_n\delta(E-E_n)$ with $A_n=(v_x^n)^2\tau_n^{\text{tr}}$. In weak disorder limit, we can replace the sum over eigenstates $n$ of the disordered Hamiltonian by integral over clean bands ($s=\pm$) and momentum $\boldsymbol{k}$: $\sum_n\to\sum_{s=\pm}\intop d\boldsymbol{k}, E_n\to E_s(\boldsymbol{k}),v_x^n\to v_{x,s}, \tau_n^{\text{tr}}\to\tau_{\text{tr}}(s,\boldsymbol{k})$. With this Eq. (\ref{eq:kubotr}) becomes
\begin{equation}
    g_{xx}=\sum_{s=\pm}\intop dE\left(-\frac{\partial f}{\partial E}\right)\intop \frac{d^2k}{(2\pi)^2}\delta(E-E_s(\boldsymbol{k}))v_{x,s}^2\tau_{\text{tr}}(s,\boldsymbol{k})
\end{equation}
After some manipulation, and using $-\frac{\partial f}{\partial E}\to \delta(E-E_F)$, we arrive at
\begin{equation}
    g_{xx}=\sum_{s=\pm}\intop \frac{d^2k}{(2\pi)^2}\delta(E_F-E_s(\boldsymbol{k}))v_{x,s}^2\tau_{\text{tr}}(s,\boldsymbol{k})
\end{equation}
Now we will use the identity $\intop \frac{d^2\boldsymbol{k}}{(2\pi)^2}\delta(E-E_F)F(\boldsymbol{k})=\oint_{E_s=E_F}\frac{dl_s(\phi)}{\nabla_{\boldsymbol{k}}E_s(\boldsymbol{k})}F(\boldsymbol{k})$ to reach 
\begin{equation}
    g_{xx} = \sum_{s=\pm}\int \frac{dl_{s}(\phi)}{(2\pi)^2|\boldsymbol{v}_s(\phi)|}v_{x,s}(\phi)^2\tau_{\text{tr}}(s, \phi)
\end{equation}
Where, $dl_s(\phi)$ is the line element along the contour $E_s(\boldsymbol{k})=E_F$, $|\boldsymbol{v}_s|$ is the magnitude of the group velocity 
\begin{equation}
   \boldsymbol{v}_s = (\lambda + s\cos{\phi}, s\sin{\phi}) 
   \label{eq:v}
\end{equation}
$\hat{k} = (\cos{\phi}, \sin{\phi})$, and $\tau_\text{tr}$ is the transport time on the Fermi surface. 

Starting from this, we will analyze how conductivity along tilt direction ($g_{xx}$) behaves as a function of tilt. For this, we divide it into three regimes: 1. Close to upright node or very small tilt ($\lambda \ll 1$) 2. Deep type II regime ($\lambda>1$) 3. Around the transition point of type I and type II, i.e., around $\lambda=1$. 

\subsection{$\lambda\ll1$}
In this regime, the Fermi surface is closed and the Fermi contour at energy $E$ for band $s$ is given by $k_s(\phi) = \frac{E}{s + \lambda\cos{\phi}}$. Here are taking $E$ small, and assume that these results will hold even when  $E\to 0$. Transport time $\tau_{\text{tr}}$ on the Fermi surface is 
\begin{equation}
    \frac{1}{\tau_{\text{tr}}(s, \phi)} = \sum_{s=\pm}\int\frac{dl_{s'}(\phi')}{(2\pi)^2|\boldsymbol{v}_{s'}(\phi')|}W(\boldsymbol{q})F_{ss'}(\theta)(1-\cos{\theta_v})
    \label{eq:tautr}
\end{equation}
Where,
$F_{ss'} = \frac{1 + ss'\cos{\theta}}{2}$ with $\theta = \phi - \phi'$ is the spinor overlap, $W(\boldsymbol{q}) = W_0e^{-q^2\xi^2/2}$ is the Gaussian disorder potential with $\boldsymbol{q} = \boldsymbol{k} - \boldsymbol{k}'$, and $\cos{\theta_v} = \frac{\lambda^2+\lambda(s\cos{\phi}+ s'\cos{\phi'}) + ss'\cos{(\phi-\phi')}}{\sqrt{1+\lambda^2+2s\lambda\cos{\phi}}\sqrt{1+\lambda^2+2s'\lambda\cos{\phi'}}}$
First we will focus on small tilt, $\lambda \ll 1$. To get tilt dependence of $g_{xx}$, let us first try to understand how does $\frac{v_{x,s}^2}{v_s}$ changes with $\lambda$. 
\begin{equation}
    \frac{v_{x,s}^2}{|v_s|} = \frac{(\lambda+s\cos{\phi})^2}{\sqrt{1+\lambda^2+ 2s\lambda\cos{\phi}}}
\end{equation}
Expanding the denominator to $O(\lambda^2)$:
\begin{equation}
    \frac{1}{\sqrt{1+\lambda^2+2s\lambda\cos{\phi}}}=1-s\lambda\cos{\phi}+\lambda^2\left(\frac{3}{2}\cos^2{\phi} -\frac{1}{2}\right) + O(\lambda^3).
\end{equation}
With this, and replacing $\cos$ terms by their angular averages, we have
\begin{equation}
\left\langle\frac{v^2_{x,s}}{|v_s|}\right\rangle=\frac{1}{2}\left[1+\frac{5}{8}\lambda^2+O(\lambda^4)\right] 
\end{equation}
Now, let us compute $\tau_{\text{tr}}$. From Eq. (\ref{eq:tautr}), we can write 
\begin{equation}
    \frac{1}{\tau_{\text{tr}}} = N(E,\lambda)\langle W(\boldsymbol{q})F(\theta)(1-\cos{\theta_v})\rangle_{\text{FS}}
\end{equation}
For $\lambda \ll 1$, we can approximate $\theta_v \simeq \theta$, $s=s'$, and $F(\theta) = \frac{1+\cos{\theta}}{2}$. This point onward We will drop the explicit band index and assume $s=+$. Expanding $k(\phi)$ in orders of $\lambda^2$,
\begin{equation}
    k(\phi) = \frac{E}{1 + \lambda\cos{\phi}}=E(1-\lambda\cos{\phi} + \lambda^2\cos^2{\phi})
\end{equation}
Approximating $k(\phi)k(\phi')$ by $\frac{k^2(\phi')+k^2(\phi)}{2}$, we can write 
\begin{equation}
    q^2(\lambda) = q^2(0)\left[1-\lambda(\cos\phi+\cos\phi') + \frac{3}{2}(\cos^2\phi + \cos^2\phi')\right]
\end{equation}
For $\lambda=0$, 
\begin{equation}
    \frac{1}{\tau_{\text{tr}}^0} = N_0(E)W_0\int_0^{2\pi}\frac{d\theta}{2\pi}e^{-a(1-\cos\theta)}\frac{1+\cos\theta}{2}(1-\cos\theta)
\end{equation}
This integral becomes $J(a)=e^{-a}\frac{I_0(a)-I_2(a)}{4}$, where $I_0(a)$ and $I_2(a)$ are modified Bessel functions of the first kind with $a = q^2\xi^2$ and $\tau_{\text{tr}}^{0}$ denotes transport time at $\lambda=0$. 

For $\lambda>0$, changes in $\tau_{\text{tr}}$ comes from $J(a_{\lambda})$. Expanding $J(a_{\lambda})$ to $O(\lambda^2)$,
\begin{equation}
    J(a_{\lambda}) = J(a) + J'(a)\delta a + \frac{1}{2}J''(a)(\delta a)^2 +..
\end{equation}
Now, since $\delta a \sim \cos \phi$, $O(\lambda)$ terms will vanish upon averaging, and our leading order term is of $O(\lambda^2)$. 
\begin{equation}
    \langle J(a_t)\rangle = J(a)\left[1-\lambda^2\left(\frac{3}{2}a\frac{J'(a)}{J(a)} + \frac{1}{2}a^2\frac{J''(a)}{J(a)}\right)\right]
\end{equation}
We now have 
\begin{equation}
    N(E,\lambda)\tau_{\text{tr}} = \frac{1}{W_0J(a)}[1+f_s(a)\lambda^2]
\end{equation}
where $f_s = \frac{3}{2}a\frac{J'(a)}{J(a)} + \frac{1}{2}a^2\frac{J''(a)}{J(a)}$
Finally, putting everything together,
\begin{equation}
    \frac{g_{xx}(\lambda)}{g_{xx}(0)}= \left[1+\frac{5}{8}\lambda^2\right]\left[1+f_s\lambda^2\right] + O(\lambda^4)=1+\left[\frac{5}{8}+f_s\right]\lambda^2 + O(\lambda^4)
\end{equation}
From this, it is clear that for small tilt the change in conductivity comes with $O(\lambda^2)$, not in linear order. This explains why conductivity does not increase much with tilt for $\lambda \ll 1$.
\subsection{$\lambda \gg1$}
Now, let us see what happens when Dirac node enters type II regime. For $\lambda > 1$, Fermi surface is no longer closed, and single Fermi surface splits into two Fermi pockets. The point $\lambda = 1$ is a non-analytic boundary, and we can not continue the $\lambda\ll1$ across it. Moreover, now since there are two Fermi pockets, there will be inter-pocket scattering. Next we will qualitatively show how $g_{xx}$ behaves with $\lambda$.

To discuss $\lambda$ dependence of $g_{xx}$, we break $g_{xx}$ into two parts: kinematic part $\frac{v_x^2}{|\boldsymbol{v}|}$, and the scattering part $\tau_{\text{tr}}$. We have seen earlier $k(\phi) = \frac{E}{u}$ with $u = 1 + \lambda\cos\phi$. In terms of $u$ and $\Delta=\lambda^2-1$, $v_x = \frac{\Delta+u}{\lambda}$ and $|\boldsymbol{v}| = \sqrt{\Delta + 2u}$. With this, the kinematic part becomes
\begin{equation}
    K = \frac{E}{(2\pi\lambda)^2}\frac{(\Delta+u)^2}{u^2\sqrt{\Delta+2u-(u-1)^2}}
\end{equation}

For the scattering part, tilt dependence mainly comes from the Gaussian factor $W(\boldsymbol{q)}$ with $q^2 \sim E^2[(\frac{1}{u}-\frac{1}{u'})^2+ \frac{2(1-\cos\theta)}{uu'}$. Now, if $u, u'$ are very different, $W(\boldsymbol{q})$ kills those contribution. We thus take $u \sim u'$, $W(\boldsymbol{q})\sim W_0e^{-a/u^2}$. For the kinematic part $K$, $\frac{(u+\Delta)^2}{u^2} = \frac{\Delta^2}{u^2} + \frac{2\Delta}{u} + 1 \equiv T_1 + T_2 + T_3$. The square root of the denominator of $K$, after expanding in $\lambda$ becomes
$\sqrt{\Delta+2u-u^2} = \lambda\sqrt{1-(\frac{u-1}{\lambda})^2}=\lambda(1+O(\lambda^{-2})$
So, including all these
\begin{equation}
    g_{xx}(\lambda^2)\sim \frac{1}{\lambda}\int_{u_0}^{1+\lambda}(T_1 + T_2 + T_3)W(u)du
\end{equation}
We will analyze each terms separately. For $T_3$, the integral converges. After some complicated integral, $T_2$ gives $\frac{\log(1+\lambda)}{\lambda}$ contribution, and finally, for $T_1$, the integral goes as $O(1/\lambda)$. Including all these, to leading order, we get
\begin{equation}
    g_{xx}(\lambda)\sim O(1/\lambda^2)
\end{equation}
So, for type II node, conductivity $g_{xx}$ decreases with $\lambda$. 
\subsection{Around $\lambda=1$}

Now let us analyze the behavior near critical tilt $\lambda = 1$. The anomalous behavior of the conductivity at the critical tilt is governed by the evolution of the Fermi surface near $\phi = \pi$, i.e. along the momentum direction opposite to the tilt. For $\lambda < 1$. Assume $\lambda = 1 - \delta$ with $0<\delta\ll1$, and expand $\phi = \pi - \psi$. With this, 
\begin{equation}
    u(\psi)=1+\lambda\cos\psi\simeq\delta+\frac{\psi^2}{2}
\end{equation}
We can write $v_x$ as $v_x = \frac{\Delta + u}{\lambda} \simeq \Delta+u$. With this, the factor multiplying $\tau_{\text{tr}}$ in conductivity is
\begin{equation}
    \frac{v_x^2}{u^2} \sim \frac{(\Delta+u)^2}{u^2}=\left(\frac{\psi^2/2 - \delta}{\psi^2/2 + \delta}\right)^2 < 1
\end{equation}
So, for $\lambda < 1$, the Fermi surface is a closed loop, and while states near 
$\phi = \pi$ are present, their contribution to transport is weakened by the velocity ratio factor that reduces their weight in the conductivity integral. At the critical tilt $\lambda=1$, the denominator $1+\lambda\cos\phi$ vanishes at 
$\phi = \pi$, so the suppression disappears and these states contribute with maximal weight. When $\lambda > 1$, the denominator of $k(\phi)$ becomes negative in a finite angular window around $\pi$, meaning that this sector no longer belongs to the Fermi surface: the closed contour splits into open pockets. 

Now, we will see the behavior of $g_{xx}$ across critical tilt $\lambda=1$ as function of $\lambda$. 
Near $\phi = \pi$  we write

\[
u(\psi) = 1 + \eta \cos(\pi-\psi)
\simeq
\begin{cases}
\delta + \dfrac{\psi^2}{2}, & \eta = 1-\delta,\\[6pt]
-\delta + \dfrac{\psi^2}{2}, & \eta = 1+\delta,
\end{cases}
\qquad 
\Delta = \eta^2 - 1 = \pm 2\delta + O(\delta^2).
\]

The velocity along tilt direction $v_x$ is
\[
v_x = \frac{1}{\eta}(\Delta+u) \simeq \Delta+u \quad (\eta\simeq 1).
\]

The conductivity integrand is
\[
\mathcal{I}_\eta(\psi)
= \frac{e^2}{(2\pi)^2}E\;
\frac{v_x^2}{u^2}\;\tau_{\rm tr}(\psi,\eta).
\]
Here we are using explicit dependence of $\tau_{\rm tr}$ on $\psi, \eta$ for ease of computation.
\begin{itemize}
\item{Critical tilt $\eta=1$}
\end{itemize}
At $\eta=1$ one has $\Delta=0$, $u=\psi^2/2$, so
\[
\frac{v_x^2}{u^2} =  1,
\]
and therefore
\[
\mathcal{I}_1(\psi) = \frac{e^2}{(2\pi)^2}E v\,\tau_c(\psi),
\]
which is finite in the vicinity of $\psi=\pi$ and $\tau_c(\psi)=\tau_{\text{tr}}(\psi,1)$.
\begin{itemize}
\item{Left of criticality $\eta=1-\delta$}
\end{itemize}
Here $\Delta=-2\delta$, and we write $u=\delta+x$. Then
\[
\frac{v_x^2}{u^2} 
= \left(\frac{x-\delta}{x+\delta}\right)^2 \equiv R_-(x,\delta)\le 1,
\]
for all $x\ge 0$ and $\delta>0$. Hence, taking $\tau_{\text{tr}}\to \tau_c$ to leading order, we have
\begin{equation}
 g_{xx}(1)- g_{xx}(1-\delta) = \frac{e^2}{(2\pi)^2}E \int d\psi \bar{\tau}_c(1-R_{-}(x,\delta))
\end{equation}
With $d\psi = dx/\sqrt{2x}$ and writing $x=\delta t$, we get
\begin{equation}
  g_{xx}(1)- g_{xx}(1-\delta) = \frac{e^2}{(2\pi)^2}E\bar{\tau}_{\text{c}} 2\sqrt{2\delta}\int_0^{\infty}\frac{\sqrt{t}}{(1+t)^2}dt = \frac{\sqrt{2}\pi}{(2\pi)^2}e^2E\bar{\tau}_{c}\sqrt{\delta} 
\end{equation}
\begin{itemize}
    \item{Right of criticality $\eta=1+\delta$}
\end{itemize}
Now, following earlier methods,
\begin{equation}
    \frac{v_x^2}{u^2}=\left(\frac{x+\delta}{x-\delta}\right)^2
\end{equation}
The integrand now becomes,
\begin{equation}
    \mathcal{I}_{1+\delta}=\mathcal{K}\left(\frac{x+\delta}{x-\delta}\right)^2\tau_{\text{tr}}(\psi,1+\delta), 
\end{equation}
Here, in contrast to the left side, the integral not over every $\phi$. For $\lambda    >1$, the integral domain excludes angles where $u(\phi)<0$ or, $\psi>\psi_0=\sqrt{2\delta}$. Thus we write
\begin{equation}
    g_{xx}(1+\delta)-g_{xx}(1) = \int_{\psi>\psi_0}(\mathcal{I}_{1+\delta}-\mathcal{I}_1)d\psi - \int_{\psi<\psi_0}\mathcal{I}_1d\psi 
\end{equation}
Near $\psi = 0$, 
\begin{equation}
    \mathcal{I}_1(\psi)=\mathcal{K}\tau(\psi)=\mathcal{K}(\tau(0)+O(\psi^2))
\end{equation}
So, the `missing-sector' contribution is
\begin{equation}
    -\int_{\psi<\psi_0}\mathcal{I}_1=-(2\sqrt{2}\mathcal{K}\tau(0))\sqrt{\delta} + O(\delta^{3/2})
\end{equation}

Now consider $\psi > \psi_0$, 
\begin{equation}
    \mathcal{I}_{1+\delta}-\mathcal{I}_1=\mathcal{K}\left\{\left[\left(\frac{x+\delta}{x-\delta}\right)^2-1\right]\tau_c(\psi) + \left(\frac{x+\delta}{x-\delta}\right)^2[\tau_{\text{tr}}(\psi;1+\delta)-\tau_c(\psi)\right\}\equiv A + B
\end{equation}
Now, we will treat each separately.
\begin{equation}
    A = \int_{\delta}^{x_{\text{max}}}\frac{dx}{\sqrt{2x}}\frac{4x\delta}{(x-\delta)^2}\tau_c(\psi(x))
\end{equation}
Following earlier methods,
\begin{equation}
    A=\frac{4}{\sqrt{2}}\sqrt{\delta}\int_1^{t_{\text{max}}}\frac{t^{1/2}}{(t-1)^2}\tau_c(\psi(t))dt
\end{equation}
This integrand becomes divergent for $t=1$. This can be regularized by both finite sized disorder and microscopic cutoff on the Dirac cone. Near the Fermi pocket edge $u=x-\delta\to0$, but the Gaussian $W(q)\propto e^{-(k_F\xi)^2/u^2}$ will suppress the divergence from $1/u$. Since $0<u<\delta$, $\frac{4x\delta}{u^2} \le \frac{8\delta^2}{u^2}$, and $d\psi=\frac{dx}{\sqrt{2x}}\le\frac{du}{\sqrt{2\delta}}$. In this range of $u$, the integral is bounded by 
\begin{equation}
    \int^{\delta}_{0}\frac{4x\delta}{u^2}\tau_cd\psi\le\frac{8\delta^2}{\sqrt{2\delta}}\tau_c^{\text{max}}(\psi)\int_{u_0}^{\delta}\frac{du}{u^2}=\frac{8\delta^2}{\sqrt{2\delta}}\tau_c^{\text{max}}(\psi)(1/\delta-1/u_0)
\end{equation} where $\tau_c(\psi)=\tau_{\text{tr}}(\psi;1)$.
The subleading order contribution is $\frac{8}{\sqrt{2}}\tau_c^{\text{max}}(\psi)\sqrt{\delta}\frac{\delta}{u_0}$. For Gaussian potential with finite $\xi$, $u_0\sim \frac{1}{\sqrt{\ln(1/\delta)}}$, and $\delta/u_0 \to 0$ as $\delta\to 0$ keeping leading order terms to $O(\sqrt{\delta)}$.

For the region $x\ge2\delta$, we can bound 
\begin{equation}
    \frac{4x\delta}{(x-\delta)^2}\le\frac{16\delta}{x}\int_{2\delta}^{x_{\text{max}}}\frac{dx}{x}\frac{1}{\sqrt{2x}}\tau_c^{\text{max}}\le \text{const.}\times \sqrt{\delta}
\end{equation}

Now let us expand $\tau_{\text{tr}}(\psi;1+\delta)=\tau_c(\psi)+O(\delta)$. We then have
\begin{equation}
    \int_{\psi>\psi_0}\left(\frac{x+\delta}{x-\delta}\right)^2[\tau_{\text{tr}}-\tau_c]d\psi=O(\delta)\int_{\psi_0}^{\psi_{\text{max}}}\left(\frac{x+\delta}{x-\delta}\right)^2d\psi
\end{equation}
Now, proceeding as earlier, we will split the integral into two two regions: $x \ge 2\delta$, and $x\in (\delta,2\delta)$. For $x \ge 2\delta$, similar calculation shows the leading order contribution is $O(\delta)$. For, the other region, it is $O(\sqrt{\delta)}$. Stiching all parts together, we get 
\begin{equation}
    g_{xx}(1+\delta)-g_{xx}(1)= -c\sqrt{\delta}
\end{equation}
where $c=\sqrt{8}\mathcal{K}\tau_c(0)$.

Since $g_{xx}(1-\delta)<g_{xx}(1)$ and 
$g_{xx}(1+\delta)<g_{xx}(1)$, the conductivity attains a 
sharp maximum at the critical tilt $\lambda=1$, and this matches perfectly with our simulation.

\twocolumngrid
\bibliographystyle{apsrev4-2}

\bibliography{thesis_bib}

\end{document}